\documentclass[pra,showpacs,preprintnumbers,tightenlines,superscriptaddress,twocolumn]{revtex4}

 \usepackage[pdftex]{graphicx}
\usepackage{dcolumn}
\usepackage{bm}
\usepackage{epsfig,color}
\usepackage{stmaryrd}
\usepackage{dsfont}
\usepackage{bbm}
\usepackage{amsmath}
\usepackage{amsfonts}
\usepackage{amssymb}
\usepackage{amscd}
\usepackage{amstext}
\usepackage{float}
\usepackage{rotating,standalone}
\usepackage{tikz}
\usepackage{soul}
\usepackage[percent]{overpic}
\usepackage{ulem}

\newcommand{\vc}[1]{\mathbf{#1}}

\newcommand{\bra}[1]{\langle\,{#1}\, |}
\newcommand{\ket}[1]{|\,{#1}\,\rangle}
\newcommand{\braket}[2]{\mbox{$\langle\,{#1}\, | \,{#2}\,\rangle$}}

\newcommand{\id}{\mathds{1}}
\newcommand{\nullop}{\ensuremath{\mathbb{O}}}
\newcommand{\sub}[2]{{#1}_{\mbox{\!\! \scriptsize #2}}}

\newcommand{\bv}[1]{\mathbf{ #1 }}

\def\beq{\begin{equation}}
\def\eeq{\end{equation}}

\newcommand{\rref}[1]{Ref.~\cite{#1}}
\newcommand{\fref}[1]{Fig.~\ref{#1}}

\newcommand{\sref}[1]{section~\ref{#1}}

\newcommand{\cref}[1]{chapter~\ref{#1}}

\newcommand{\Cref}[1]{Chapter~\ref{#1}}

\newcommand{\aref}[1]{appendix~\ref{#1}}
\newcommand{\bref}[1]{(\ref{#1})}

\newcommand{\op}[1]{\ensuremath{\hat{#1}}}
\newcommand{\im}{\ensuremath{\mathrm{i}}}

\newcommand{\symbvecspace}{\ensuremath{V}}
\newcommand{\symbspinspace}{\ensuremath{S}}
\newcommand{\stylecs}[1]{\ensuremath{\mathfrak{#1}}}

\newcommand{\spaces}{\ensuremath{\symbvecspace}}
\newcommand{\spacel}{\ensuremath{\mathcal{\symbvecspace}}}

\newcommand{\spaceallspins}{\ensuremath{\mathcal{\symbspinspace}}}
\newcommand{\spacepinsmaller}[1]{\ensuremath{\mathcal{\symbspinspace}^{(#1)}}}

\newcommand{\spacefull}{\ensuremath{\stylecs{\symbvecspace}}}
\newcommand{\spacefullj}[1]{\ensuremath{\stylecs{\symbvecspace}_{#1}}}

\newcommand{\textinfig}[1]{\tikz [baseline] \node {#1};}

\newcommand{\basiss}{\ensuremath{B[\spaces]}}
\newcommand{\basisl}{\ensuremath{B[\spacel]}}

\newcommand{\basissallspins}{\ensuremath{B[\spaceallspins]}}
\newcommand{\basisspinsmaller}[1]{\ensuremath{B[\mathcal{\symbspinspace}^{(#1)}]}}

\newcommand{\basissfullj}[1]{\ensuremath{B[\stylecs{\symbvecspace}_{#1}]}}

\newcommand{\Eddmax}{\ensuremath{E_{\mathcal{DD}}^{\mathrm{max}}}}
\newcommand{\Rmin}{\ensuremath{R_{\mathrm{min}}}}

\newcommand{\ncolumnwidth}{0.9\columnwidth}
\begin{document}

\title{Orthogonal flexible Rydberg aggregates}
\author{K.~Leonhardt, S.~W\"uster and J.~M.~Rost }
\affiliation{Max Planck Institute for the Physics of Complex Systems, N\"othnitzer Strasse 38, 01187 Dresden, Germany}
\email{karlo@pks.mpg.de}
\begin{abstract}
We study the link between atomic motion and exciton transport in flexible Rydberg aggregates, assemblies of highly excited light alkali atoms, for which motion due to dipole-dipole interaction becomes relevant. In two one-dimensional atom chains crossing at a right angle adiabatic exciton transport is affected by a conical intersection of excitonic energy surfaces, which induces controllable non-adiabatic effects. 
A joint exciton/motion pulse that is initially governed by a single energy surface is coherently split into two modes after crossing the intersection. The modes induce strongly different atomic motion, leading to clear signatures of non-adiabatic effects in atomic density profiles. We have shown how this scenario can be exploited as an exciton switch, controlling direction and coherence properties of the joint pulse on the second of the chains [K.~Leonhardt {\it et al.}, Phys.~Rev.~Lett. {\bf 113} 223001 (2014)].
In this article we discuss the underlying complex dynamics in detail, characterise the switch and derive our isotropic interaction model from a realistic anisotropic one with the addition of a magnetic bias field.
\end{abstract}
\pacs{
32.80.Ee,  
82.20.Rp,  
34.20.Cf,    
31.50.Gh   
}
\maketitle

\section{Introduction}
%
Over the last decade  fundamental phenomena in many-body Rydberg systems  have been discovered, from dipole blockade~\cite{lukin:quantuminfo,urban:twoatomblock,gaetan:twoatomblock,tong:blockade} and antiblockade~\cite{cenap:manybody,singer:blockade} over long-range molecules~\cite{Greene:LongRangeMols,liu:ultra_long_range_2009} to excitonic dynamics due to resonant dipole-dipole interactions~\cite{cenap:motion,wuester:cradle,moebius:cradle,wuester:CI,zoubi:VdWagg,moebius:bobbels,barredo:trimeragg,guenter:EITexpt,cenap:emergent,muelken:excitontransfer}.
 Underlying all of these are strong long range interactions \cite{park:dipdipbroadening,park:dipdipionization,li_gallagher:dipdipexcit,westermann:transfer,ravets:foersterdipdip}, which also make Rydberg atoms promising for quantum computing~\cite{Saffman-M&oslash;lmer-Quantuminformationwith-2010}, quantum simulators~\cite{Weimer-Buchler-Rydbergquantumsimulator-2010,schempp:spintransport} and model systems for biological processes~\cite{muelken:exciton:survival,schoenleber:immag}.
\newline
It also became apparent, that the interplay of atomic motion and excitonic dynamics~\cite{frenkel_exciton} yields entanglement transport~\cite{wuester:cradle,moebius:cradle} and conical intersections (CIs)~\cite{wuester:CI}. The latter play a crucial role in many quantum chemical processes, where they enable highly non-adiabatic dynamics on an ultrafast timescale~\cite{dantus_femotchemistry}. They may also protect the DNA structure from damage by UV radiation~\cite{perun:dna_protectionbyCI}. In \rref{leonhardt:switch} we have explored
how conical intersections affect the wavepackets of atomic motion, which describe the entanglement transport and the associated energy and momentum transfer~\cite{wuester:cradle}. We have shown that the CI can split those wavepackets, resulting in a superposition of two different excitation transfer processes. Since the relative weight of the processes after the splitting can be tuned through the system geometry and hence CI position, this gives rise to a sensitive switch for exciton transport properties.
\newline
 Here we show the evolution of the underlying exciton spectra as a function of time in more detail. We also provide a parameter space survey for the operation of the conical intersection switch. Our results are derived from a model of isotropic dipole-dipole interactions. We finally show qualitatively similar features using 
 the inherently anisotropic dipole-dipole interactions, where the anisotropy is suppressed through the application of an external magnetic field. 
\newline
Dipole-dipole forces between atoms in two different Rydberg states $\ket{s}$ and $\ket{p}$ cause a strong interdependence of atomic motion and Rydberg state dynamics~\cite{cenap:motion}. On a homogeneous chain
of all but one atom in state $\ket{s}$, a single $\ket{p}$ excitation will delocalize forming a Frenkel exciton~\cite{frenkel_exciton}. If the chain has a dislocation formed by two more closely spaced atoms, the exciton localises on these and in a repulsive state causes a combined pulse of chain dislocation, excitation and entanglement to propagate through the chain, akin to Newton's cradle~\cite{wuester:cradle}. In the scenario considered here, this combined pulse is directed towards a second chain orthogonal to the first, causing atoms to reach a configuration with a conical intersection in the exciton energy spectrum.  The intersection causes non-adiabatic effects, triggering two different modes of pulse propagation on the second chain. The basic mechanism by which the conical intersection acts as a switch between these modes, can best be understood considering the essential subunit of two orthogonal atomic dimers.
\newline
The paper is organized as follows: In \sref{basics}, we discuss Rydberg aggregates and their numerical treatment. We then proceed in \sref{sec:nges4} to describe a double Rydberg dimer, with two atoms freely moving on each of two orthogonal chains. We study the consequences of a conical intersection and characterise the dynamics on each of the two participating potential surfaces. Subsequently, in \sref{sec:nges8} we extend this setup to a seven atom system, with three and four atoms on the two chains. In \sref{exciton_switch} we highlight how the conical intersection can be functionalised as a switch controlling the transport dynamics on the second chain. Finally in \sref{Bfield} we examine the feasibility to realise our isotropic interaction model experimentally. The appendix contains technical details regarding the engineering of isotropic interactions and removal of spin degrees of freedom using a magnetic field.

\section{Rydberg aggregates}
 \label{model}

\label{basics}
\begin{figure}[tb]
\centering
\epsfig{file={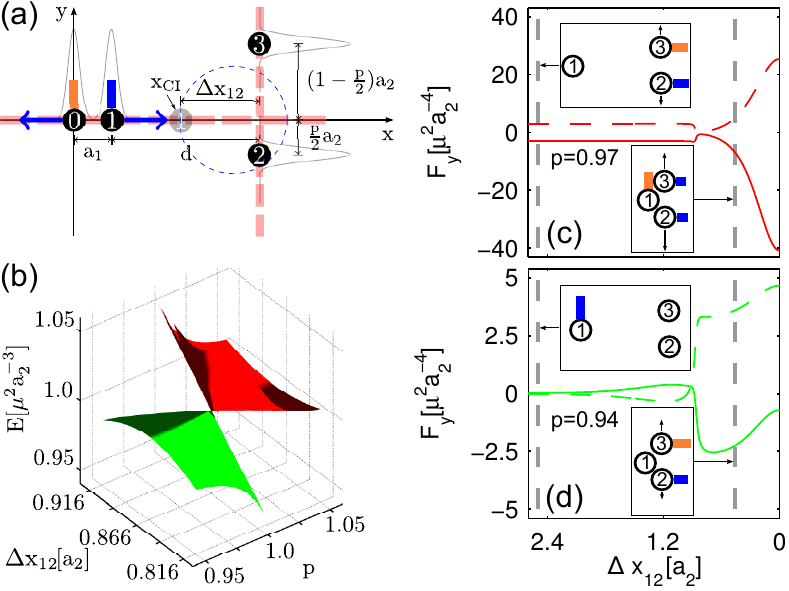},width=\ncolumnwidth}
\caption{(color online) (a) Orthogonal atom chains with one Rydberg dimer each. Atoms 0 and 1 initially share an excitation. Due to the ensuing repulsion (blue arrows) atom 1 reaches the conical intersection at $\sub{x}{CI}$. The colored bars visualise the excitation amplitude on each atom $c_n = \braket{\pi_n}{\sub{\varphi}{rep}}$, with orange $c_n>0$ and blue $c_n<0$. The origin of the coordinate system is set to the mean initial position of atom~0. (b)~The repulsive energy surface $U_{\mathrm{rep}}$ (red) and adjacent surface $U_{\mathrm{mid}}$ (green) of the trimer sub-unit (atom~1, 2 and 3) near the CI.
(c,d) Forces on atom~2 (solid lines) and atom~3 (dashed lines), for the repulsive surface (red,c) and adjacent surface (green,d). The insets show atomic positions and the excitation distribution $c_n$ of exciton states and forces for the indicated values of~$\Delta x_{12}$, which denotes the distance between atom~1 and the vertical chain. The parameter $p$ controls the degree of symmetry of the trimer, where $p=1$ corresponds to an isosceles trimer configuration. Figure reproduced from~\cite{leonhardt:switch}.
\label{system_sketch_4}}
\end{figure}

We study a system of $N$ Rydberg atoms with masses $M$, all with the same principal quantum number $\nu$  which we restrict for the sake of clarity to the two cases $\nu=44$ and $\nu=80$. With $N_x$ of the atoms  constrained on the $x$-axis, and $N_y$ on the $y$-axis their total number is $N=N_x+N_y$, for a simple example see \fref{system_sketch_4}a.
All atoms can  move freely in only one dimension, with their positions described by the 
 vector $\vc{R}~=~(\vc{R}_1, \dots, \vc{R}_n)^{\rm{T}}$. The one-dimensional confinement could for example be realized by running laser fields and optical trapping of alkali Rydberg atoms~\cite{Li:kuzmich:atomlighentangle}, or earth alkali Rydberg atoms through their second valence electron~\cite{rick:Rydberglattice}.
Furthermore, we assume atoms prepared such that only  one atom is in an angular momentum $p$ state, all the other atoms are  in angular momentum $s$ states.
This allows us to expand the electronic wavefunction in the single excitation basis $\{\ket{\pi_k}\}$, where $\ket{\pi_k}=\ket{s\dots p\dots s}$
denotes a state with the $k$th atom in the $p$ state~\cite{cenap:motion,wuester:cradle}. 

\subsection{Rydberg-Rydberg interaction and the electronic Hamiltonian}
Interaction potentials between Rydberg atoms can be determined by diagonalizing a dimer Hamiltonian in a restricted electronic state space, using the dipole-dipole approximation \cite{book:gallagher}. Here we capture the essential features of these potentials into an effective model, including van-der-Waals (vdW) interactions between two atoms in the same state ($ss$ or $pp$), and resonant dipole-dipole interactions between two atoms in different states ($sp$), leading to the electronic Hamiltonian %
\begin{equation}
 \op{H}^{\rm{el}}(\vc{R})=\op{H}_{\rm{dd}}(\vc{R}) + \op{H}_{\rm{vdw}}(\vc{R}),
  \label{eq:elechamiltonian}
\end{equation}
with  resonant dipole-dipole interaction Hamiltonian 
\begin{subequations}
\label{eq:interactions}
\begin{equation}
 \op{H}_{\rm{dd}}(\vc{R})=-\mu^2\sum_{\substack{m,n=1;\\ m\neq n}}^{N}R_{mn}^{-3}\ket{\pi_m}\bra{\pi_n}
  \label{eq:elechamiltonian-dd}
\end{equation}
and non-resonant van-der-Waals interaction
\begin{equation}
 \op{H}_{\rm{vdw}}(\vc{R})=-\id\sum_{\substack{m,n = 1;\\ m\neq n}}^N \frac{C_6}{2R_{mn}^6}\,,
  \label{eq:elechamiltonian-vdw}
\end{equation}
\end{subequations}
where $\id$ is the unit operator in the electronic space
and $R_{mn} = |\vc R_m- \vc R_n|$  the distance between atoms $m$ and $n$. The interactions \bref{eq:elechamiltonian-dd} are isotropic, determined by  the scaled radial matrix element  $\mu=d_{\nu,1;\nu,0}/\sqrt{6}$. In
  \sref{Bfield} and \aref{isotrope_dip_dip_int} we discuss how this simplification can be realized using a magnetic field and isolating specific azimuthal angular momentum states. 
Using just one coefficient $C_6 < 0$ in \eqref{eq:elechamiltonian-vdw} ensures repulsive behavior at very short distances for all electronic states, assuming identical vdW interactions between Rydberg atoms in $s$ and $p$ states for simplicity. Their difference in reality can give rise to interesting effects at shorter distances \cite{zoubi:VdWagg} that are, however, not relevant in our context.

Diagonalizing the electronic Hamiltonian for fixed nuclei,
\begin{equation}
 \op{H}^{\rm el}(\vc{R})\ket{\varphi_k(\vc{R})}= U_k(\vc{R})\ket{\varphi_k(\vc{R})}\,,
\label{eq:eveqn}
\end{equation}
gives the eigenstates $\ket{\varphi_k(\vc{R})}$ called Frenkel excitons~\cite{frenkel_exciton}
and the eigenenergies $U_k(\vc{R})$ which form Born-Oppenheimer surfaces (BO surfaces).

\subsection{Initial state}
\label{initial_state}
The initial state of the system can be written as a direct product
\begin{equation}\label{initialPsi}
\ket{\Psi^{\rm ini}}=\ket{\psi_{\rm el}^{\rm ini}}\otimes\ket{\chi_{\rm nuc}^{\rm ini}},
\end{equation}
where $\ket{\psi_{\rm el}^{\rm ini}}$ is the initial electronic state and $\ket{\chi_{\rm nuc}^{\rm ini}}$ 
the initial spatial position state. We choose $\ket{\chi_{\rm nuc}^{\rm ini}}$ such that the initial position space representation for $N$ atoms,
\begin{subequations}
\begin{align}
\label{gaussian}
&\chi_{\rm nuc}^{\rm ini}(\vc R)=\braket{\vc R}{\chi_{\rm nuc}^{\rm ini}},
\\
&=\left(2\pi\sigma^2\right)^{-\frac{N}{4}}\prod_n^N \exp{  \left(-\frac{|\bv{R}_{n} - \bv{R}^{(0)}_n | ^2}{4 \sigma^2}  \right)},
\label{inipos}
\end{align}
\end{subequations}
is a product of Gaussians, centered on the atoms initial positions $\bv{R}^{(0)}_n$. We enforce $\bv{R}_{n}=x_n \bv{e}_x$ for atoms on the horizonal chain ($n\leq N_x$) and $\bv{R}_{n}=y_n \bv{e}_y$ for atoms on the vertical chain ($n> N_x$), where $\bv{e}_{x,y}$ are unit vectors in the $x$, $y$ directions. This makes sure that the atoms have no initial position uncertainty in directions orthogonal to their chain. The longitudinal width $\sigma$ models an experimental preparation in the lowest oscillator state of a harmonic trap.

\subsection{Propagation scheme for atomic motion}
\label{prop_scheme}
For larger numbers of atoms, solving the time dependent Schr\"odinger equation  of the full 
Hamiltonian
\begin{equation}
 \op{H}(\vc{R}) = -\frac{\nabla^2_{\vc{R}}}{2M} + \op{H}^{\rm{el}}(\vc{R}).
\label{eq:hamiltonian}
\end{equation}
 for the motion in $\vc R$ is not feasible in a reasonable time.
 Instead,  we take the Wigner transform of  the atomic Gaussian wave function \bref{gaussian} as a weighting distribution for initial conditions $\gamma_{i}=\{\vc P_{i},\vc R_{i}\}$ to propagate classical trajectories  fulfilling  Newton's equation
\begin{equation}
 M\ddot{\vc{R}}=-\nabla_{\vc{R}}U_s(\vc{R}).
\label{eq:newton_equation}
\end{equation}
The mechanical force exerted on  the atoms is derived  from \emph{a single} Born-Oppenheimer surface $U_s$ of the electronic Hamiltonian. However, we do allow
 transitions to other adiabatic surfaces $U_{k}$ during the time evolution according to 
 Tully's fewest switching algorithm \cite{tully:hopping2,tully:hopping:veloadjust,barbatti:review_tully}.
 These jumps, are random but occur with the probability 
for a transition between two surfaces $n$ and $m$, which is proportional to the non-adiabatic coupling vector
\begin{equation}
 \vc{d}_{mn}(\vc{R})=\bra{\varphi_m(\vc{R})}\nabla_{\vc{R}}\ket{\varphi_n(\vc{R})}.
 \label{nonadiabcoupl}
\end{equation}
Simultaneously, we propagate the electronic wave function of the Rydberg aggregate  $\Phi(\vc R(t)) = \sum_{n}c_n(t)\ket{\pi_n}$, expressed in the atomic (diabatic) basis $\{|\pi_{n}\rangle\}$, by solving the 
matrix equations for the coefficients $c_{n}(t)$,
\begin{equation}
\im \hbar \dot{c}_k(t)= \sum_{\ell}\hat H^{\rm el}_{k\ell}(t)\, c_\ell(t)\,,  
\label{eq:propagation_diabatic_states}
\end{equation}
where
\begin{equation}
 \begin{split}
\hat H^{\rm el}_{k\ell}(t) &\equiv \langle\pi_{k}|H^{\rm el}(\vc{R}(t))|\pi_{\ell}\rangle\\
 &= -\frac{\mu^{2}}{R_{k\ell}^{3}(t)} - \left(\sum_{\substack{m,n = 1;\\ m\neq n}}^N  \frac{C_{6}}{2R_{mn}^{6}(t)}\right)\delta_{k\ell}.
\end{split}
\label{eq:HamiltonianMEs}
\end{equation}
In addition to coefficients in the diabatic basis, we will also refer to those in the adiabatic basis defined by 
 $\Phi(\vc R(t)) = \sum_{n}\tilde{c}_n(t)\ket{\varphi_n(\vc{R}(t))}$.
Each trajectory is a self consistent solution of the coupled atomic and electronic equations of motion, 
\bref{eq:newton_equation} and \bref{eq:propagation_diabatic_states}
taking into account in addition the random jump sequence $s$ among the BO surfaces  discussed above. 
 With a recorded jump-sequence  $s_{i}(t)$, each trajectory $i$  represents a realization.
More precisely, the propagation sequence for a single trajectory with initial conditions $\gamma_{i}$ at time $t=0$ starting on the BO surface $U_{s_{i}(0)}$ consists of the following steps:
\begin{enumerate}
\item The electronic Hamiltonian is diagonalized and at any given time a single eigenenergy $U_{s(t)}(\vc R(t))$  is the  potential which exerts  force on the atoms. \label{enumeration:diag}
\item The atomic positions are propagated with \bref{eq:newton_equation} and the electronic states   with \bref{eq:propagation_diabatic_states} for one time step $\Delta t$.
\item We determine whether the surface index $s_i$ undergoes a stochastic jump after this time step to $s_{i}(t+\Delta t)\neq s_i(t)$ (see \cite{cenap:motion} for the precise prescription).
\item The new positions $\vc R(t+\Delta t)$ lead to new eigen states and -energies. Hence, we repeat the procedure from point \ref{enumeration:diag} with  updated time variable $t \leftarrow t+\Delta t$.
\end{enumerate}
 We use a single realization $s_{i}$ for each initial condition $\gamma_{i}$, which is in practice sufficient to converge the stochastic character of the jumps between surfaces if a large number of trajectories is propagated whose initial conditions randomly sample the initial Wigner distribution. Results obtained in this way agree very well with full quantum results for three atoms where quantum calculations can be done \cite{wuester:cradle,moebius:cradle,moebius:bobbels,leonhardt:switch}.
Since Tully's algorithm is expected to be an even better approximation for  a larger number of dimension we confidently  have adapted it to   our present problem as described above.

\section{Nonadiabatic dynamics}
\label{nonad_dynamics}
Equipped with the tools to describe dynamics where the aggregate can make transitions between different adiabatic surfaces/ excitons, we would like to control these transitions in order to steer exciton transfer
in a complex landscape of potential surfaces. It is well known that conical intersections -- where adiabatic surfaces touch each other --  lead to enhanced non-adiabatic transfer from on surface to another. Previously, we investigated transport of excitation and entanglement 
in linear flexible Rydberg chains without non-adiabatic transfer \cite{wuester:cradle,moebius:cradle},
and the opposite, non-adiabatic processes near conical intersections without transport in ring configurations~\cite{wuester:CI}. 

A system exhibiting both features, transport and induced non-adiabatic transfer, is a T-shape configuration of two one-dimensional chains of atoms. 
We will present their dynamics with three and four atoms respectively on the two chains, using Li atoms with a mass of $M=11000$~au. First, however, we investigate the simplest possible realization, a double dimer with two atoms on each of two orthogonal lines as sketched in \fref{system_sketch_4}.

\subsection{Two perpendicular dimers}
\label{sec:nges4}
 We use Rydberg states in principal quantum number $\nu=44$, leading to a transition dipole moment of $\mu=1000$~au. For simplicity we set $C_6=0$ in this section. The parameters of the initial atomic configuration were set to $a_1=2.16\ \rm{\mu m}$, $a_2=5.25\ \rm{\mu m}$ and $d=8.5\ \rm{\mu m}$. To sample the initial nuclear (Wigner) distribution
with Gaussian width of $\sigma=0.5\ \rm{\mu m}$ about each atomic position \bref{gaussian}, we have propagated $10^5$ trajectories.

A  single p-excitation in the system shared between atom~0 and 1 leads to their mutual repulsion as indicated by the blue arrows in \fref{system_sketch_4}a and  enables 
 atom~1 to move towards the vertical chain. On this way the three atoms~1-3 form a triangular sub-unit
 corresponding to the \emph{ring trimer} studied in \cite{wuester:CI}.
 The CI of the trimer is realised for $p=1$, $d=\sqrt{3}/2 a_2$ in \fref{system_sketch_4}a, where the three atoms form an equilateral triangle. To illustrate the CI, we show in \fref{system_sketch_4}b the two intersecting energy surfaces as a function of two selected atomic position variables. The upper surface, shaded in red, will be hereafter referred to as the repulsive surface $\sub{U}{rep}$ (with exciton state $\ket{\sub{\varphi}{rep}}$), as it always entails repulsive interactions of nearby atoms. The lower surface at the intersection, shaded in green, will be referred to as the adjacent surface $\sub{U}{adj}$ (with exciton state $\ket{\sub{\varphi}{adj}}$ \cite{footnote:Umidadj}). Further surfaces are not shown and play no significant role.
 
We will now systematically construct and interpret the atomic motion triggered by the initial excitation, firstly by analyzing typical trajectories and their energy spectra, then by constructing the atomic densities on the repulsive and adjacent adiabatic surface, which finally will enable us to understand the full evolution of the atomic density. We consider its evolution in time, spatially resolved, in terms of population of adiabatic surfaces and regarding the purity of the state.

\subsubsection{Evolution and energy spectra of typical trajectories}
\label{sec:doubledimer:typicaltrajs}
   
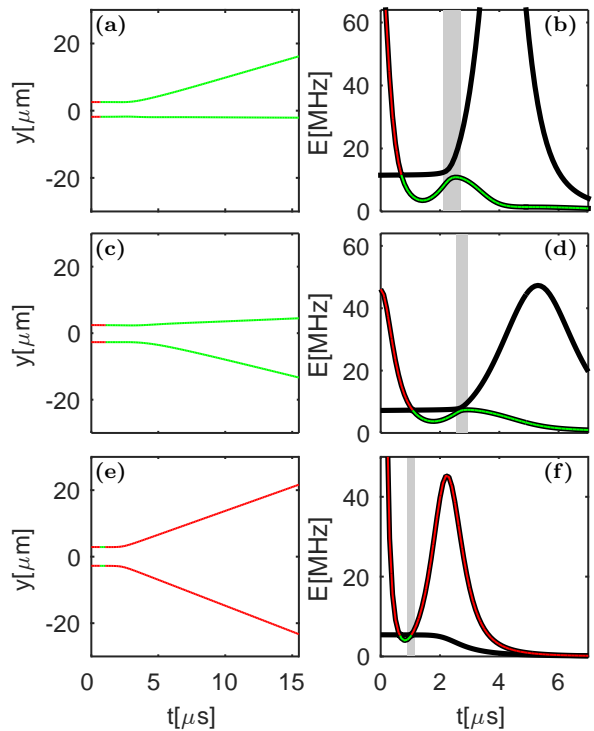
\begin{figure}[htb]
\centering
\begin{overpic}[width=\ncolumnwidth,tics=2]{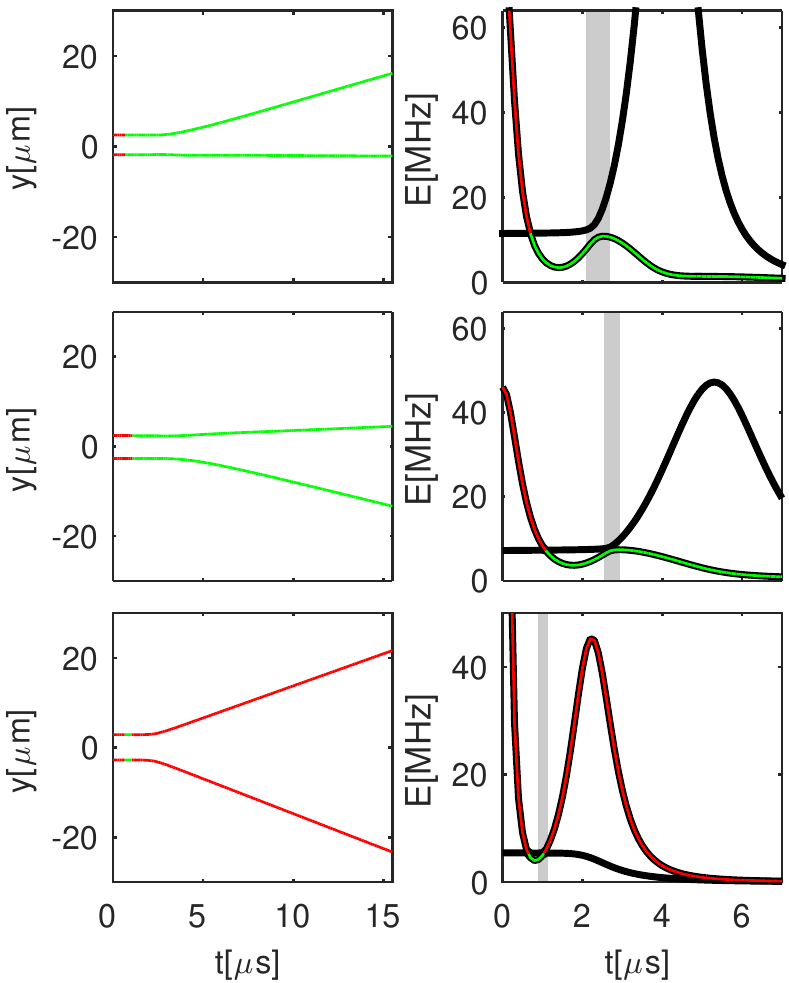}
 \put (10.8,96.7) {\textinfig{\footnotesize\textbf{\color{black}{(a)}}}}
 \put (10.8,66) {\textinfig{\footnotesize\textbf{\color{black}{(c)}}}}
 \put (10.8,35.4) {\textinfig{\footnotesize\textbf{\color{black}{(e)}}}}
 \put (72.3,96.7) {\textinfig{\footnotesize\textbf{\color{black}{(b)}}}}
 \put (72.3,66)   {\textinfig{\footnotesize\textbf{\color{black}{(d)}}}}
 \put (72.3,35.4) {\textinfig{\footnotesize\textbf{\color{black}{(f)}}}}
\end{overpic}
\caption{\label{fig:Nges4_single_trajectories}
Selected single trajectories $\vc R(t)$ of atoms~2 and~3 in~(a,c,e) with corresponding time resolved energy spectra~(black lines) and potential energy $U_{s(t)}(\vc R(t))$~(colored line) in~(b,d,f).
For both, trajectories and potential energy, evolution on the repulsive surface is marked with red lines and evolution on the adjacent surface with green lines.
(a,c) Two characteristic trajectories that stay on the adjacent surface. The two trajectories are almost mirror symmetric around the $x$-axis.
(e) A characteristic trajectory jumping back to the repulsive surface after passing the vicinity of the CI.
The grey area marks the vicinity of the CI between the adjacent and the repulsive surfaces, ruling the dynamics. 
The earlier crossings are less relevant for our dynamics here, see~\cite{discussion:CI_crossing}.
}
\end{figure}

Since we neglect all uncertainties \emph{transverse} to each of the trapping beams (see \bref{gaussian}),  atoms zero and one have well defined $y$-coordinates. However, the atoms on the vertical line have a distribution 
in $y$ about their central position such that  different trajectories have
different asymmetry parameters $p$. We distinguish the ``symmetric'' part of the nuclear wavepacket   with $p\approx1$ (a typical trajectory is shown in \fref{fig:Nges4_single_trajectories}e) from the  ``asymmetric'' part which realizes other $p$ values (two trajectories that are nearly related through vertical mirroring are shown 
 in \fref{fig:Nges4_single_trajectories}a,c). One can see that initially, symmetric and asymmetric trajectories evolve very similarly. In particular, both of them jump from the repulsive surface (red) to the one we have called the adjacent surface of the trimer (green) a bit before one $\mu s$ into the dynamics, see the energy spectra \fref{fig:Nges4_single_trajectories}b,d,f. This point in parameter space represents a further conical intersection with only trivial consequences for our dynamics here, see \cite{discussion:CI_crossing}. During this initial time, atom 0 and 1 simply repel (not shown), until atom 1 sets 2 and 3 into motion around $2 \mu$s.
Shortly afterwards (during the grey shaded time interval) the dynamics of the trajectories is ruled by the essential conical intersection of the trimer.
Here, the symmetric part, coming near the degenerate point of the CI, is more likely to make a transition to the repulsive surface than
the asymmetric part, which stays dominantly on the adjacent surface and passes the CI at a greater distance, see \fref{system_sketch_4}b). This is reflected in the energy spectrum of the trajectories by a very narrow avoided crossing (it appears actually as a crossing within the resolution of \fref{system_sketch_4}f) for the symmetric trajectory in the grey shaded area. 
 Consequently, the symmetric trajectory follows a diabatic path, jumping from the adjacent surface to the repulsive one, while the two asymmetric trajectories that miss the CI move adiabatically in their energy landscape \fref{fig:Nges4_single_trajectories}b,d), with a relevant avoided crossing in the grey shaded area that is wider allowing them to stay on the adjacent surface. A final transition to the repulsive surface 
leads to a significant separation of atoms 2 and 3 in the course of time, while the asymmetric trajectories that stay on the adjacent surface contain one atom remaining almost at rest,
while the other one, initially at {\em larger} distance to the horizontal chain experiences the stronger force (\fref{system_sketch_4}c) and moves away
\cite{footnote:asymmforce}.

 Exciton spectra such as shown in \fref{fig:Nges4_single_trajectories} should be observable with micro-wave spectroscopy of Rydberg aggregates, such as in \cite{park:dipdipbroadening,celistrino_teixeira:microwavespec_motion}.

\subsubsection{Atomic densities on the adiabatic surfaces forming the conical intersection}

\begin{figure}[htb]
\centering
\begin{overpic}[width=\ncolumnwidth,tics=2]{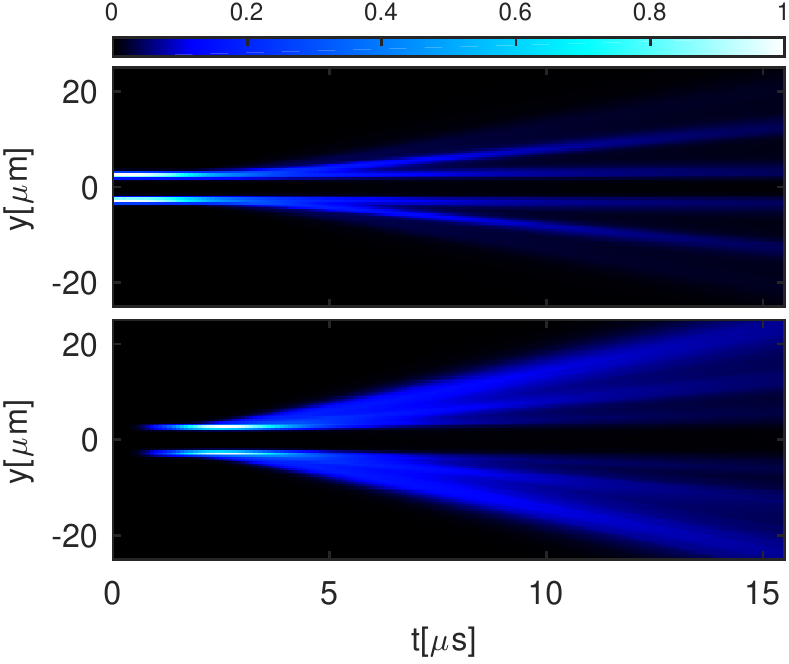}
 \put (14.1,73)  {\textinfig{\footnotesize\textbf{\color{white}{(a)}}}}
 \put (14.1,40.8) {\textinfig{\footnotesize\textbf{\color{white}{(b)}}}}
\end{overpic}
\caption{\label{fig:Nges4_partial_densities_chain2}
Atomic densities of atoms~2 and~3
on the adjacent surface (a),  and on the repulsive surface (b).
Atomic positions were binned into one or the other according to the trajectory surface index $s$. The maximum value of the data over both (a,b) is set to unity with a common normalisation.}
\end{figure}

To evolve the Rydberg aggregate from its initial (Wigner)-distribution, we have calculated $i = 1,...,\sub{N}{traj} = 10^{5}$ trajectories, the surface index $s_{i}(t)$ of which at each instant of time allows us to show the atomic density \cite{footnote:density}  on the adjacent and repulsive surface separately as done in \fref{fig:Nges4_partial_densities_chain2} for atoms 2 and 3. One directly recognises that the wave packet enters the CI on the adjacent surface
(no density in \fref{fig:Nges4_partial_densities_chain2}b for short times) and that through the CI roughly half of the density is transferred to the repulsive surface where the two repelling branches show a wide distribution. On the adjacent surface, on the other hand, the four-fold branching is due to the asymmetric character of its individual trajectories as illustrated in \fref{fig:Nges4_single_trajectories}a,c). 
%

\subsubsection{Time dependent observables of the double dimer}

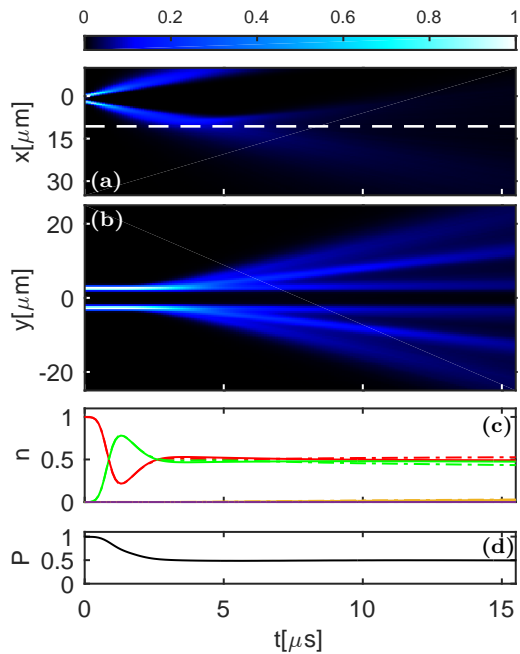
\begin{figure}[htb]
\centering
\begin{overpic}[width=\ncolumnwidth,tics=2]{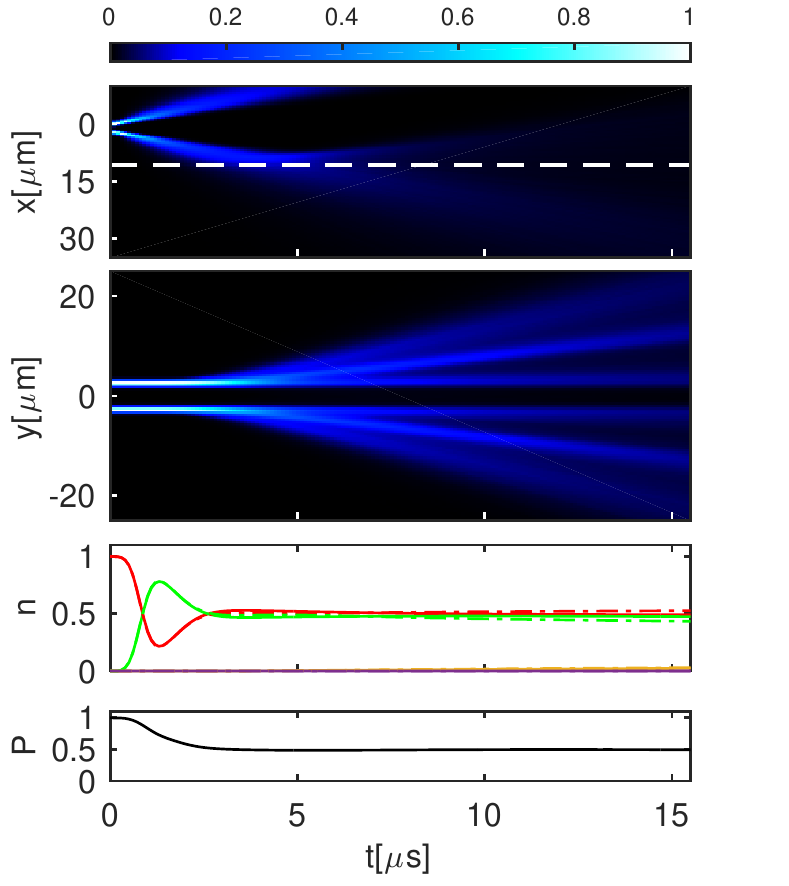}
 \put (12,73)  {\textinfig{\footnotesize\textbf{\color{white}{(a)}}}}
 \put (12,67.) {\textinfig{\footnotesize\textbf{\color{white}{(b)}}}}
 \put (71.5,35.7) {\textinfig{\footnotesize\textbf{\color{black}{(c)}}}}
  \put (71.5,17.2) {\textinfig{\footnotesize\textbf{\color{black}{(d)}}}}
\end{overpic}
\caption{\label{fig:Nges4_total_densities_pops_purity}
Dynamics of atomic motion and electronic populations; total atomic density of atom~0 and 1  in (a) and
atom~2 and~3 in (b).
(c)~Adiabatic populations, $\overline{|\tilde{c}_n|^2}$, (solid lines) and trajectory fractions $f_n=\sum_i^{\sub{N}{traj}} \delta_{n,s_i(t)}/\sub{N}{traj}$ (dashed dottes lines) on the repulsive surface (red) and adjacent surface (green). Here $\overline{\cdots}$ denotes a trajectory average. (d) Purity $P=\rm{Tr[}\hat{\sigma}^2\rm{]}$ of the reduced electronic density matrix~\cite{wuester:cradle,footnote:purity}. All results are averaged over $10^5$ realizations. The maximum value of the data in (a,b) is individually set to one.
}
\end{figure}

We finally present the full time evolution of the double dimer in \fref{fig:Nges4_total_densities_pops_purity}, where we clearly recognize in (a) the initial repulsion of atoms 0 and 1 of the horizontal chain. Furthermore, we can appreciate and understand the complicated fanning out of density in the vertical chain after the CI as a superposition of the wide two-branch distribution on the repulsive surface and the narrower four finger double fork
on the adjacent surface, \fref{fig:Nges4_total_densities_pops_purity}b. The population on both surfaces as a function of time in \fref{fig:Nges4_total_densities_pops_purity}c confirms that the wave packet starts out on the repulsive surface (red) and continues after the first crossing on the adjacent surface (green), while it is split at the conical intersection to finally populate both surfaces roughly equally. As a consequence, the purity  \cite{footnote:purity}  of the wave packet decreases from unity to one half in \fref{fig:Nges4_total_densities_pops_purity}d.

\subsection{The seven atom T-shape system}
\label{sec:nges8}
The double dimer was a necessary prerequisite to create a coherent exciton pulse along a   path defined by  atoms positioned in space:
We now know that the atoms on the vertical chain will not  attract each other after the atomic wave packet has passed the CI crossing, which is the first condition to continue excitation transport from the horizontal onto the vertical chain. The transport from its initiation, by creating a repulsive exciton localised on the first two atoms,  to the CI on the horizontal chain 
follows the dynamics we have already established with ``Newton's cradle'' in~\rref{wuester:cradle}. However, for more atoms in the vertical chain, the CI may induce features which qualitatively go beyond what we have seen in the double dimer. Hence we extend our system to three atoms in the horizontal and four atoms in the vertical chain, keeping the configuration of two orthogonal lines as shown in \fref{system_sketch_7}. The resulting setup corresponds to 
two linear atomic "Newton's Cradle" configurations as in~\rref{wuester:cradle} that intersect at a right angle, such that transport involves a CI.

\subsubsection{Atomic configuration}
\label{sec:nges8_geom}
\begin{figure}[htb]
\centering
\epsfig{file={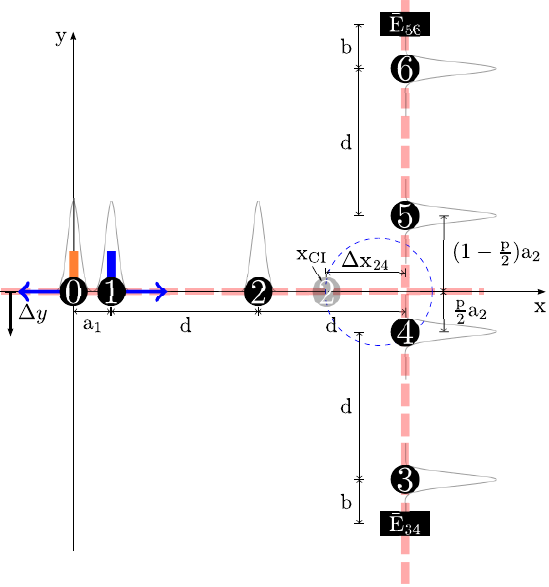},width=\ncolumnwidth} 
\caption{\label{system_sketch_7}
Sketch of a seven atom system combining features of adiabatic entanglement transport and CI dynamics. 
Three atoms are placed on the horizontal line and four on the vertical one. Atom~0 and 1 are prepared in a repulsive exciton at the initial time $t_0$, resulting in the excitation distribution shown by orange and blue bars, as described in the caption of \fref{system_sketch_4}. Atom 2 comes close to a CI configuration, when it is at position $\sub{x}{CI}$. Around that moment, the excitation almost exclusively resides on atoms~2, 4 and 5, since all other atoms are far away from this trimer. Atom 4 and 5 are accelerated at this time on the two different surfaces, already described in the four atom section. We analyse whether or not the atom pair $(3,4)$ $[(5,6)]$ finally contains an exciton-motion pulse propagating in the negative (positive) $y-$direction. This is quantified by the binary entanglement \cite{wuester:cradle} $E_{34}$ [$E_{56}$] at the moment that atom $3$ [$5$] reaches the location indicated by a black box, termed "entanglement readout". We finally use the indicated displacement ${\Delta}y$ of the horizontal chain to steer entanglement transport upwards or downwards.
}
\end{figure}
We place the atoms initially as shown in~\fref{system_sketch_7} and prepare the system in the repulsive exciton localized on atoms~0 and 1. As before, this leads to a repulsive motion, such that atom~1 kicks atom~2, which eventually transfers its energy and electronic excitation to atom 2 as familiar from Newton's cradle. Once atom~2 reaches $x_{\mathrm{CI}}$ the system can again be understood by considering the atomic trimer now formed by atoms~2, 4 and 5.

\subsubsection{Fully coherent exciton transport to the vertical chain}
\label{reptransport}

The simplest case essentially involves only a single BO surface.
For sufficient, negative $\Delta y$, atom 2 misses the CI and traverses the y-axis much closer to atom 5 than to atom 4. In this case   Newton's cradle like binary exciton transfer from atom 0 to 1 over 1 to 2 on the horizontal chain will carry over to upward transport on the vertical chain
via transfer from atom 2 to 5 and ultimately 5 to 6, realising fully adiabatic dynamics with respect to the energy surfaces. Analogously one gets for positive $\Delta y$ ultimate exciton transfer to atoms 4 and 3. Hence changes in $\Delta y$ switch between downwards or upwards exciton transport, as discussed further in \sref{exciton_switch}.

Less clear is what happens with the excitation and entanglement transport if atom 2 approaches the vertical chain 
with $\Delta y \approx 0$. As we know from the double dimer, motion of
atoms~4 {\em and} 5 is then affected by the CI through dynamics on its constituting energy surfaces, the adjacent and the repulsive one.
To elucidate what this means for the ensuing excitation transport on the vertical chain we present calculations
with Rydberg excitation $\nu=80$, which gives a transition dipole moment of $\mu=3371\ \rm{au}$ of the lithium atoms. The  parameters of the perpendicular chains are $a_1=6\ \rm{\mu m}$, $a_2=9.5~\rm{\mu m}$,  $d=22\ \rm{\mu m}$ and we use the same initial Gaussian distributions \bref{gaussian} for the atomic positions.

 \subsubsection{Simultaneous dynamics on several surfaces}
\begin{figure}[htb]
\centering
\begin{overpic}[width=\ncolumnwidth,tics=2]{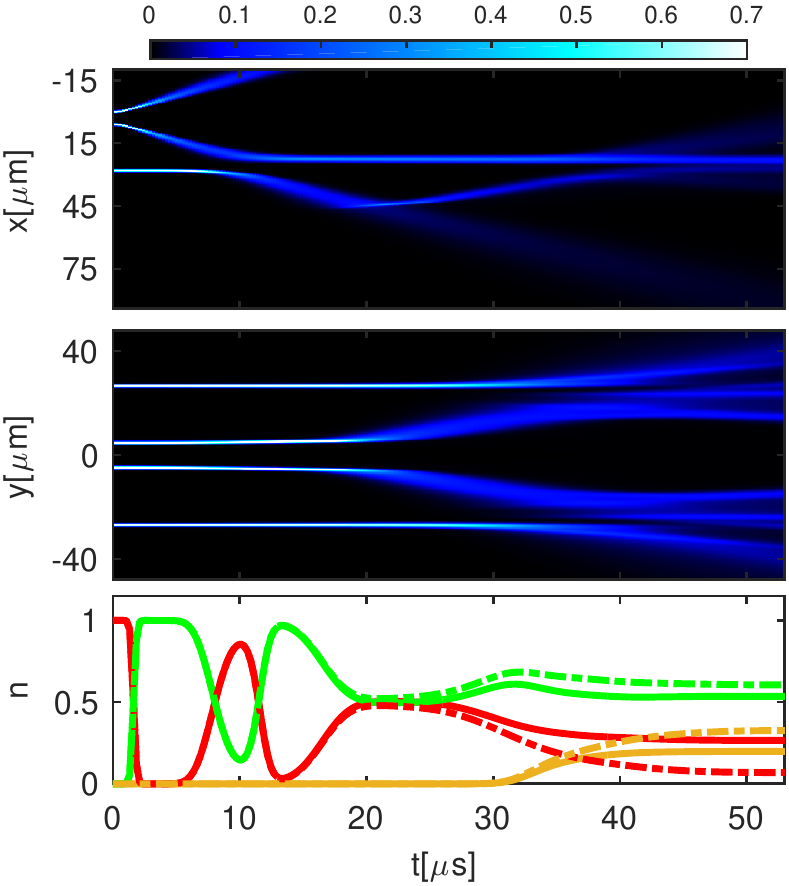}
 \put (13,67)  {\textinfig{\footnotesize\textbf{\color{white}{(a)}}}}
 \put (13,60.5) {\textinfig{\footnotesize\textbf{\color{white}{(b)}}}}
 \put (81.5,29.5) {\textinfig{\footnotesize\textbf{\color{black}{(c)}}}}
\end{overpic}
\caption{\label{fig:Nges7_total_densities_pops}
Dynamics of atomic motion and populations. (a) Total atomic density of atom~0 to~2.
(b) Total atomic density of atom~3 to~6.
(c) Adiabatic populations (solid lines) and fractions (dashed dottes lines) of repulsive surface (red), second most energetic (adjacent) surface (green) and third most energetic surface (yellow).
All data is averaged over 1~million realizations. The maximum value of the data in (a,b) is individually set to one. To highlight details at lower densities, all values between 0.7 and 1 are represented with white.
}
\end{figure}
We first discuss the case 
${\Delta}y=0\ \rm{\mu m}$ implying that the incoming atomic wave packet directly hits the CI. 

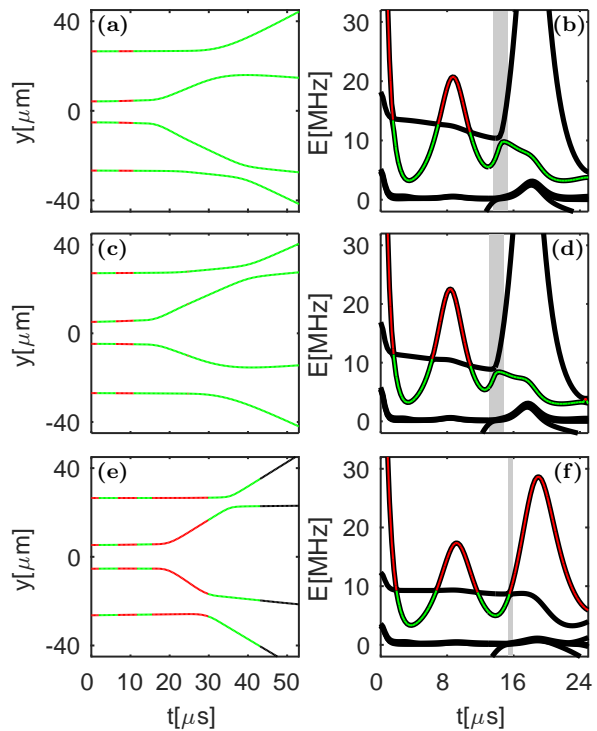
\begin{figure}[htb]
\centering
\begin{overpic}[width=\ncolumnwidth,tics=2]{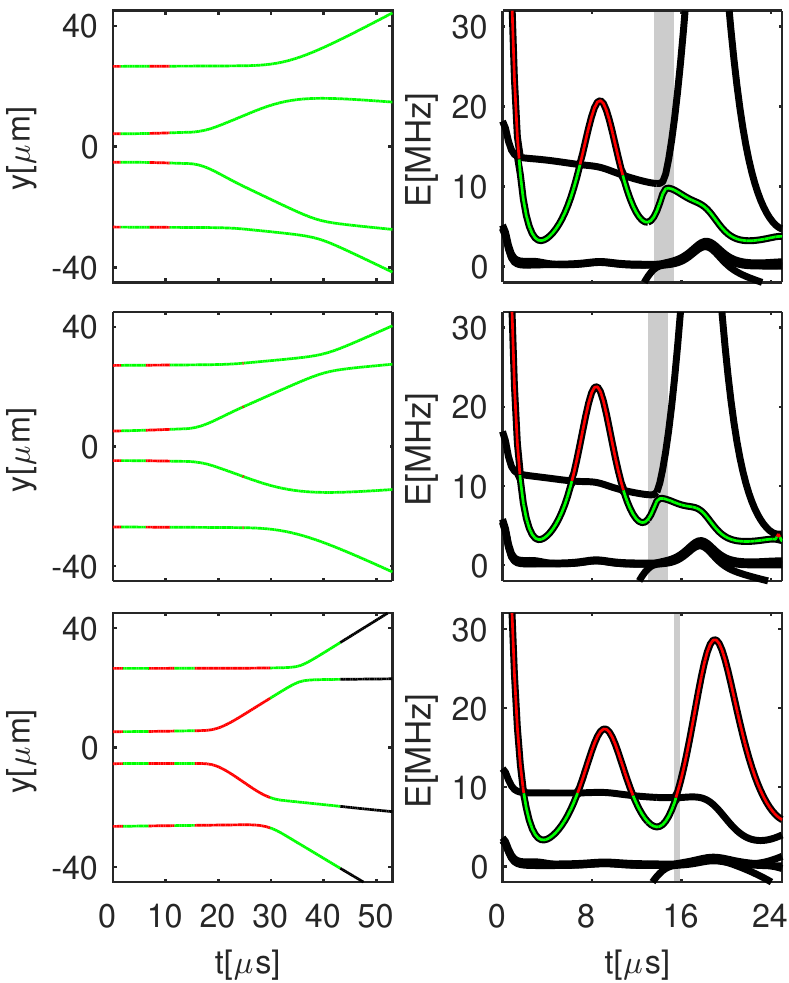}
 \put (11.1,96.7) {\textinfig{\footnotesize\textbf{\color{black}{(a)}}}}
 \put (11.1,66) {\textinfig{\footnotesize\textbf{\color{black}{(c)}}}}
 \put (11.1,35.4) {\textinfig{\footnotesize\textbf{\color{black}{(e)}}}}
 \put (73.5,96.7) {\textinfig{\footnotesize\textbf{\color{black}{(b)}}}}
 \put (73.3,66) {\textinfig{\footnotesize\textbf{\color{black}{(d)}}}}
 \put (73.6,35.4) {\textinfig{\footnotesize\textbf{\color{black}{(f)}}}}
\end{overpic}
\caption{\label{fig:Nges7_single_trajectories}
Atomic positions from single trajectories for atom~3-6 (a,c,e) with corresponding eigenenergies in (b,d,f) for $\Delta y=0~\mu$m, $a_2=9.5~\mu$m. (a,c)~Two different trajectories ending up on the adjacent surface. (e)~A trajectory returning to the repulsive surface after being in the vicinity of the CI. (b,d,f) Time resolved energy spectra (black lines) and potential energy (colored line), for position trajectories on the left. For both, positions and potential energy, evolution on the repulsive surface is marked with red and evolution on the adjacent surface with green.
The trajectories shown in (e) evolve in the end on a third surface, marked with black lines.
The gray area marks the CI and its vicinity. For a discussion of the earlier curve crossings see~\cite{discussion:CI_crossing}.
}
\end{figure}
The total atomic density on the horizontal chain  \fref{fig:Nges7_total_densities_pops}a clearly shows the repulsive behavior of the atoms transferring the initial momentum and excitation shared by atoms 0 and 1 from atom~1 to atom~2. Once atom 2 approaches
the vertical chain it enters the region of the CI which leads,  as for the double dimer, to a distribution of the atomic density
on the vertical chain over the repulsive and adjacent surface   
 (\fref{fig:Nges7_total_densities_pops}b). The exciton spectra of selected single trajectories on these surfaces shown in \fref{fig:Nges7_single_trajectories}
 behave quite similarly as for the double dimer. In the beginning on the horizontal chain the trajectories switch diabatically a couple of times  between the adjacent and repulsive surface
 before their paths finally become qualitatively different close to the CI (grey shaded region) when the two asymmetric trajectories shown in \fref{fig:Nges7_single_trajectories}a,b,c,d stay on the adjacent surface, which then rules their subsequent dynamics on the vertical chain, while the more symmetric trajectory 
 of \fref{fig:Nges7_single_trajectories}e,f jumps to the repulsive surface. The main difference compared to \fref{fig:Nges4_total_densities_pops_purity} that one recognises is the transfer of momentum to the two new outer atoms 3 and 6 on the vertical chain.

 \subsubsection{Inversion of excitation transport direction on the adjacent surface}
 \label{inversion}
 
\begin{figure}[htb]
\centering
\begin{overpic}[width=\ncolumnwidth,tics=2]{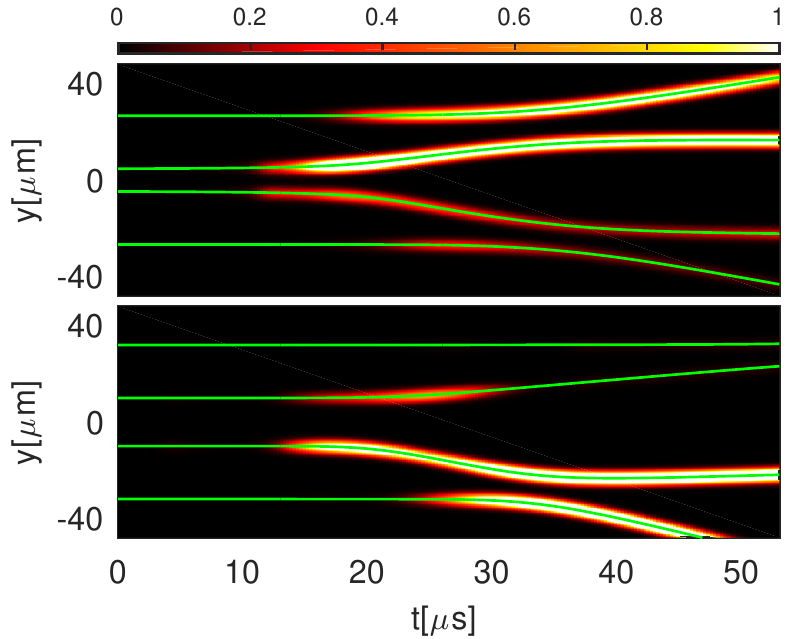}
 \put (14.1,71)  {\textinfig{\footnotesize\textbf{\color{white}{(a)}}}}
 \put (14.1,40) {\textinfig{\footnotesize\textbf{\color{white}{(b)}}}}
\end{overpic}
\caption{\label{fig:Nges7_partial_densities_chain2}Qualitatively different dynamics on the repulsive and adjacent surface.
Mean atomic positions (green lines), with $p$-excitation probability of the combined repulsive and adjacent surfaces (black-red-yellow-white shades) on each atom of the vertical chain, ${\Delta}y=1.5{\mu}$m (a) Case that mostly remains on the adjacent surface, with $a_2=9.5{\mu}$m. (b) Dominantly repulsive case with $a_2=20{\mu}$m. The excitation probability for atom $n$ is represented by Gaussians of a selected fixed width, normalized to $\sum_n |\braket{\pi_n}{\sub{\varphi}{n}(\bv{R})}|^2$ (where $\sum_n$ runs over the repulsive- and adjacent surfaces) and centered on the mean position of atom $n$. 
}
\end{figure}

Based on the understanding gained in \sref{sec:doubledimer:typicaltrajs}, we can now understand an essential qualitative difference between transport dynamics on the repulsive and adjacent surfaces. To this end, we visualize in \fref{fig:Nges7_partial_densities_chain2} the allocation of excitation transport onto the different atoms of the vertical chain, for two cases that differ only in the parameter $a_2$. While in both cases, atom $2$ passes most closely to atom $4$,  the repulsive surface realizes the intuitively expected transport continuing with atom $4$, while the adjacent surface instead leads to transport continued with atom $5$ in \emph{the opposite direction}.
As explained in \sref{sec:doubledimer:typicaltrajs}, this can be directly traced back to differences between the repulsive and adjacent exciton states near the essential conical intersection.
%

\section{Exciton switch}
\label{exciton_switch}

Depending on how the passage through or near the CI distributes population onto the repulsive or adjacent energy surface, either of the transport mechanisms discussed above can be made dominant. This population distribution depends on the width and position of the multi-dimensional atomic wavepacket describing all co-ordinates $\bv{R}$, as well as the effective size of the CI which is essentially controlled by the corresponding multi-dimensional velocity $\dot{\bv{R}}$ \cite{wuester:CI}. All these features can be varied as a function of our configuration parameters $a_1$,  $a_2$, $d$ and ${\Delta}y$. 

\begin{figure}[htb]
\centering
\epsfig{file={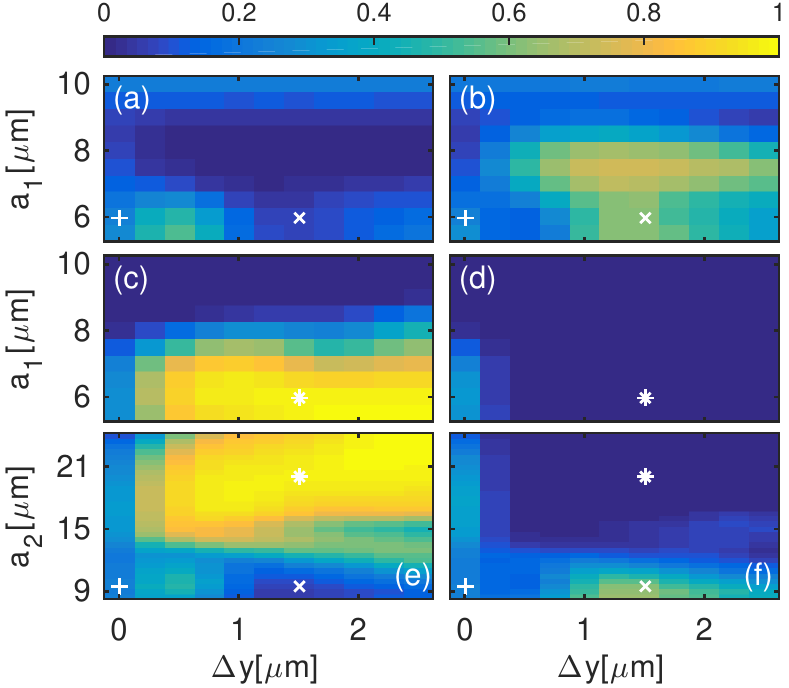},width=\ncolumnwidth}
\caption{\label{switch_parameterspace}
(color online) Response of the exciton switch to control parameters. We show the bipartite entanglement transported downwards in $y$, $\bar{E}_{34}$ (left column) and upwards, $\bar{E}_{56}$ (right column). Parameters are $\nu=80$ and $d=22~\mu$m. Both entanglement readouts (see \fref{system_sketch_7}) are placed in a distance of $b=0.3d$ from atom~3 and~6 respectively, on the vertical chain. (a-d) Entanglement as a function of $a_1$ and $\Delta y$, for $a_2=9.5~\mu$m in (a-b) and $a_2=20~\mu$m in (c-d). (e-f) Entanglement as a function of $a_2$ and $\Delta y$, for fixed $a_1=6~\mu$m.
The markers ($\ast$, $\times$, $+$) highlight extreme cases already presented in \cite{leonhardt:switch}:
($\ast$) high entanglement transport to atoms~(3,4),
($\times$) high entanglement transport to atoms~(5,6),
($+$) equal entanglement transport towards atoms~(3,4) and atoms~(5,6).
}
\end{figure}

Extending the work of \cite{leonhardt:switch} we show a parameter space survey of the ensuing "exciton switch" in \fref{switch_parameterspace}. To characterize the exciton transport in a given configuration, we use the binary entanglement 
of the final atoms on the vertical chain, at the moment when the first of them reaches its "entanglement readout" location, described in \fref{system_sketch_7}. For a definition of the binary electronic entanglement $\bar{E}_{ij}$ of two neighboring atoms $i,j$, see \cite{wuester:cradle}. A large value of $\bar{E}_{34}$ [$\bar{E}_{56}$] corresponds to a coherent exciton-motion pulse travelling in the downward (upward) direction. 

For various combinations of $a_1$,  $a_2$, and ${\Delta}y$, we plot $\bar{E}_{34}$  and $\bar{E}_{56}$ in \fref{switch_parameterspace}, determined numerically from simulations as in \fref{fig:Nges7_total_densities_pops} with a reduced number of trajectories. The three cases selected in \cite{leonhardt:switch} are shown as symbols ($\ast$, $\times$ and $+$). To understand the inversion of the entanglement transport direction for varying $a_2$ at fixed $\Delta y$, please refer to our discussion in \sref{inversion}.
%

\section{Isotropic versus anisotropic dipole-dipole model}
\label{Bfield}

The model of interactions \bref{eq:elechamiltonian} employed so far was isotropic, neglecting the angular dependence of dipole-dipole interactions \cite{book:gallagher,noordam:interactions}, and the presence of several azimuthal sub-levels of the Rydberg $\ket{p}$ state. In general we have to expand our basis to $\ket{\pi_n,m_l}=\ket{s\dots (p,m_l)\dots s}$, where $m_l$ is the magnetic quantum number of atom $n$ in the $\ket{p}$ state. For a one-dimensional atomic chain, the model \bref{eq:elechamiltonian} with basis $\ket{\pi_n}$ can be recovered by restricting the initial state to $m_l=0$, when the quantisation axis is parallel to the atomic chain \cite{moebius:cradle}. Dipole-dipole interactions then do not populate any other azimuthal states.

For the phenomena discussed in this article, we require $\sub{\hat{V}}{dd}(\bv{R})<0$ at all $\bv{R}$, for the system to display the conical intersection between the \emph{repulsive} and adjacent surface. Since additionally atomic distances here span two spatial dimensions, the procedure to eliminate the azimuthal degree of freedom described above is no longer possible. To ensure the correct sign of interactions in this case, we concentrate on the azimuthal sub-levels $m_l=1$ with quantisation axis perpendicular to the $x$-$y$ plane. To prevent undesired coupling to the $m_l=-1$ states, we additionally impose a uniform external magnetic field $\bv{B}=B_z\bv{\hat{z}}$, where $\hat{z}$ is a unit vector in the $z$-direction. This creates an energy offset ${\Delta}E = m_l \mu_B B_z$ for all electronic states, effectively decoupling those with $m_l=-1$ from $m_l=1$. We discuss formally in \aref{isotrope_dip_dip_int} how this results in our isotropic model \bref{eq:elechamiltonian} in the limit of an infinitely strong magnetic field, and how corrections to this model for realistic field strengths can be calculated. These corrections leave the qualitative results unchanged as we will show.

\begin{figure}[htb]
\centering
\begin{overpic}[width=\ncolumnwidth,tics=2]{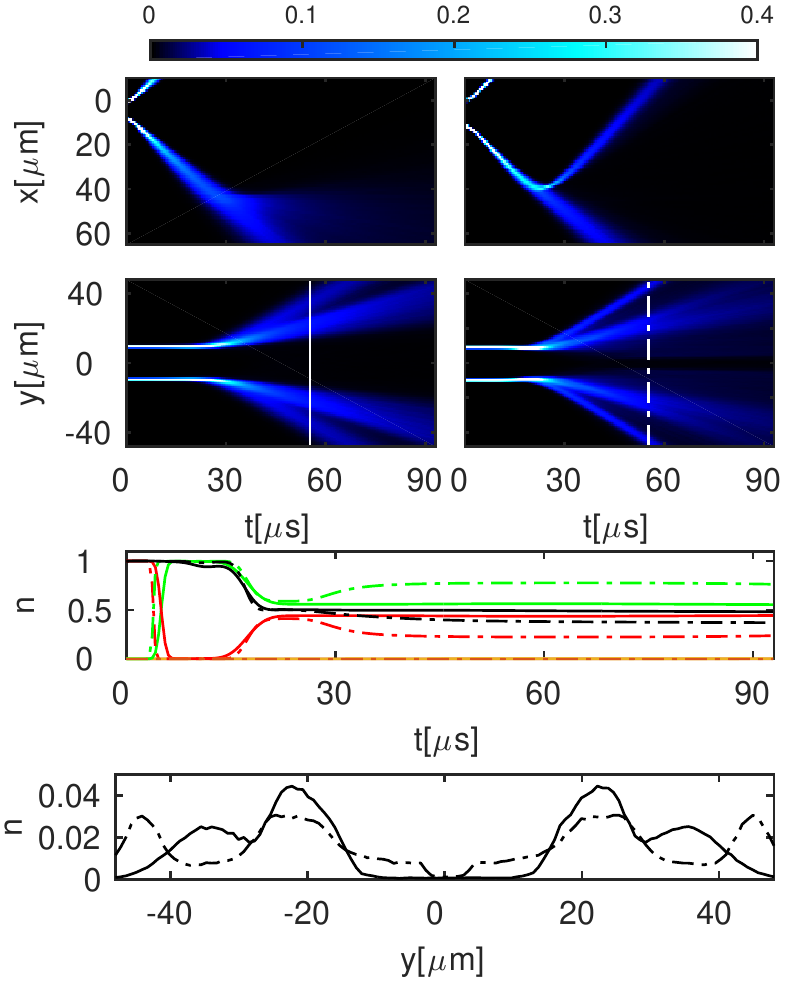}
 \put (12.5,77) {\textinfig{\footnotesize\textbf{\color{white}{(a)}}}}
 \put (12.5,69.7) {\textinfig{\footnotesize\textbf{\color{white}{(c)}}}}
 \put (47,77) {\textinfig{\footnotesize\textbf{\color{white}{(b)}}}}
 \put (47,69.7) {\textinfig{\footnotesize\textbf{\color{white}{(d)}}}}
 \put (72.5,41.6) {\textinfig{\footnotesize\textbf{\color{black}{(e)}}}}
 \put (72.5,19) {\textinfig{\footnotesize\textbf{\color{black}{(f)}}}}
\end{overpic}
\caption{\label{Bfield_doubledimer}
(color online) For the case of \fref{fig:Nges4_total_densities_pops_purity}, we compare our model of \sref{model} with one additionally allowing different azimuthal sub-levels $m_l$, anisotropic interactions and a magnetic field, see \aref{isotrope_dip_dip_int}. In both models we used the parameters $\nu=80$, $a_2=19~\mu$m and $d=40~\mu$m.
To obtain qualitatively the same dynamics in both, $a_1$ is adjusted individually, compensating quantitative differences in potentials. The anisotropic model uses $a_1=8~\mu$m and the isotropic one $a_1=11.8~\mu$m. (a,c) Atomic density of atom~0 and atom~1. (b,d) Atomic density of atom~2 and 3, where (a,c) are for the anisotropic model and (b,d) for the isotropic one.
To highlight details at lower densities, all value between 0.4 and 1 are represented with white.
 (e) Adiabatic populations on the (red line) repulsive- and (green line) adjacent energy surface and purity (black lines). We compare the anisotropic (solid line) with the isotropic model (dash-dotted line). (f) Cut through the atomic densities at the time indicated by white lines in (b,d) for both models, with line-styles as in (e).}
\end{figure}

To this end we now repeat the simulations of \fref{fig:Nges4_total_densities_pops_purity} using a model that allows all magnetic states for all four atoms in the presence of an external magnetic field, described in \aref{isotrope_dip_dip_int} and further in \cite{leonhardt:3dswitch}. The motion of the atoms is still constrained onto two orthogonal lines, as in the rest of the article. We demonstrate in \fref{Bfield_doubledimer} that qualitatively the same main features are found as for the isotropic model. However, for the greatest resemblance the parameters of both models have to be chosen slightly different due to the quantitative difference of potentials, which affect most importantly the initial acceleration of atoms $0$ and $1$, in turn controlling the relative population of the two energy surfaces after CI crossing, seen in panel (e). For \fref{Bfield_doubledimer} we have adjusted $a_1$ in both models separately to achieve a rough $50-50$ splitting on the two surfaces.
Both variants then qualitatively agree, in particular regarding clear signatures of multiple populated Born Oppenheimer surfaces in the snapshot shown in panel (f). 

%
\section{Conclusions}
\label{conclusions}

We have considered assemblies of dipole-dipole interacting Rydberg systems, whose motion and quantum state dynamics is affected by a conical intersection.
For atoms free to move on either of two orthogonal chains, we have shown that the CI in the spectrum acts as a selector, through which parts of the wave packet where atoms are spaced symmetrically around one of the chains tend to be transferred to the repulsive surface, while the rest remains on the adjacent surface. As a result of the splitting onto two surfaces, the total atomic density shows several characteristic features, clearly signaling the crossing of a conical intersection. 

The CI subunit of the two chains, with four atoms in two orthogonal dimers, can be understood as a building block for more complex systems. The block allows branching and switching of dislocation pulses within the system. Since the fraction of total population on either of the surfaces sensitively depends on the motion velocity and the width of the nuclear wave packet, the switching can be tuned or externally controlled. We have shown that this can be achieved through small changes of the confinement geometry, extending the work of \cite{leonhardt:switch} by a detailed parameter space survey of the ensuing exciton switch. Possible future extensions could be 
 velocity adjustments of Rydberg atoms via external fields, e.g.~\cite{breeden:rydberg_stark_acceleration}.

In this article we have employed an isotropic interaction model with just one participating azimuthal quantum number, which results in numerical simplifications and easier interpretation of quantum dynamics. We have demonstrated how this model can experimentally be realized using a magnetic field. However, the essential physics reported, the splitting of a Rydberg atom wave-packet onto multiple Born-Oppenheimer surfaces through acceleration by resonant dipole-dipole interactions is a generic feature of higher dimensional Rydberg aggregates. It neither relies on any of the above simplifications (as shown here) nor on the confinement of atomic motion on one-dimensional lines (as we will report in \cite{leonhardt:3dswitch}).

Finally, note that similar transport processes on different energy scales could be studied through Rydberg dressed dipole-dipole interactions \cite{wuester:dressing,genkin:dressedbobbles,moebius:cat}, or could rely on Rydberg atoms immersed in host cold atom clouds \cite{moebius:bobbels,wuester:cannon}  instead of individual atoms.

\acknowledgments
We gratefully acknowledge fruitful discussions with Alexander Eisfeld and Sebastian M{\"o}bius. 

\appendix

\section{Isotropic dipole-dipole interactions}
\label{isotrope_dip_dip_int}

As demonstrated in \sref{Bfield}, an external magnetic field can be used to reduce the number of electronic angular momentum states participating in the dynamics of Rydberg aggregates and thus render the interactions predominantly isotropic.
Here we describe the underlying model and analytically derive an expression for the resulting effective interactions, including anisotropic correction terms. We additionally discuss spin of the Rydberg atoms, and show that the same mechanism that renders interactions isotropic also removes spin as a degree of freedom.

\begin{figure}[htb]
\centering
\epsfig{file={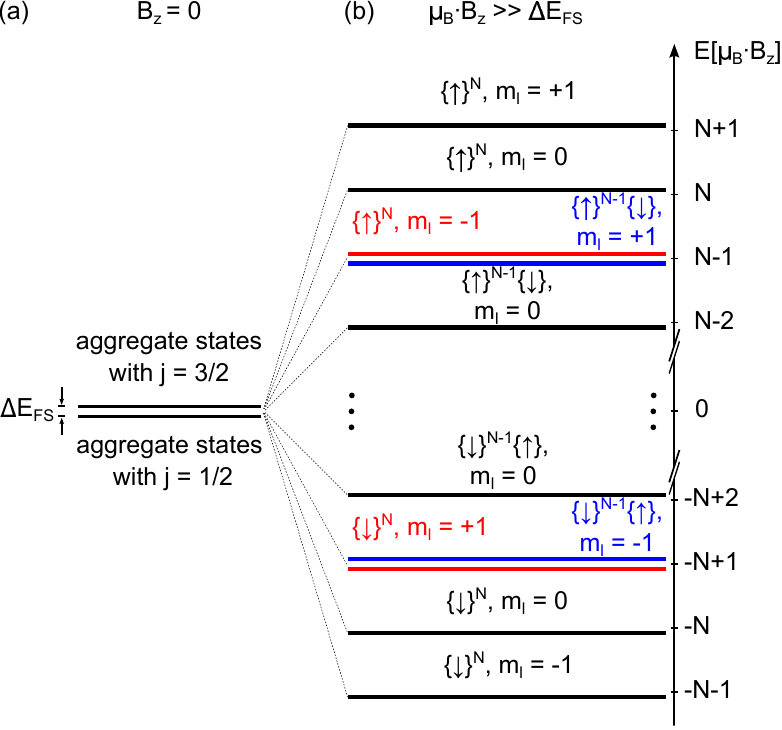},width=\ncolumnwidth} 
\caption{\label{fig:magnetic_field:energy_levels}
Sketch of energy splittings for the aggregate states.
(a) Without magnetic field, the finestructure separates aggregate states where the p-states have total angular momentum $j=1/2$ from the ones with $j=3/2$.
(b) An external magnetic field within the strong field regime, where $\mu_B \cdot B_z >> \Delta E_{FS}$, splits the aggregate states into different manifolds with energy gaps $\mu_B \cdot B_z$ for $(\Delta M_S,\Delta m_l) = (0, \pm 1)$ and energy gaps of the order of $\Delta E_{FS}$, for $(\Delta M_S,\Delta m_l) = (\pm 1, \mp2)$ (energy gap between neighboring blue and red lines).
The notation $\{\uparrow\}^{N_\uparrow}\{\downarrow\}^{N_\downarrow}$ is a short form for a spin configuration with $N_\uparrow$ spins orientied upwards ($m_s=1/2$) and $N_\downarrow$ spins oriented downwards ($m_s=-1/2$). 
}
\end{figure}

Besides including the azimuthal quantum number in our extended single excitation basis $\ket{\pi_n,m_l}$, see \sref{Bfield}, we also have to describe the spin of the atoms, where it suffices to label the z-component for each atom $i$ by $m_s^{(i)}$.
We start by defining the Hamiltonian for the interaction of the magnetic field with $N$ atoms, given by
\begin{equation}
 \hat{H}_{\mathcal{MF}}(B_z) = \mu_\mathrm{B}B_z\sum_{i=1}^{N}\hat{L}_{z}^{(i)} + 2\hat{S}_{z}^{(i)}.
\label{eq:H_MF}
\end{equation}
The magnetic field $B_z$ points in z-direction and $\hat{L}_{z}^{(i)}$, $\hat{S}_{z}^{(i)}$ denote the operators of angular momentum and spin in z-direction for the $i$th atom, and $\mu_\mathrm{B}$ is the Bohr magneton.
A spin configuration for the aggregate is uniquely defined by the tuple $\vc{M}_S:=[m_{s}^{(1)}\dots m_{s}^{(N)}]^\mathrm{T}$. We denote the corresponding state with $\ket{\vc{M}_S}:=\ket{m_s^{(1)}}_{1}\dots\ket{m_s^{(N)}}_{N}$, which is the product state of all single atom spin states, labeled for the $i$th atom with $\ket{m_s^{(i)}}_{i}$.
Introducing the quantum number for the z-component of the 
aggregate spin
\begin{equation}
 M_S(\vc{M}_S) := \sum_{i=1}^N m_s^{(i)},
\label{eq:M_S_def}
\end{equation}
which is the sum over all individual spin quantum numbers, the energy shift of the magnetic field is for $(M_S, m_l)$ states given by 
\begin{equation}
\Delta E_{\mathcal{MF}}(B_z,M_S,m_l) = \mu_\mathrm{B}B_z (2 M_S + m_l).
\label{eq:delta_E_MF}
\end{equation}
The detuning between the $m_l=1$ and $m_l=-1$ states inside a single $M_S$-manifold is $2\mu_\mathrm{B} B_z$. This has to be larger than the anisotropic matrix elements of dipole-dipole interactions, which for 
Lithium atoms and our typical parameters also makes it larger than the finestructure-splitting. This implies we are in a strong field regime, where spin and angular momentum couple separately to the magnetic field, which effectively removes the finestructure and gives an energy level structure sketched in \fref{fig:magnetic_field:energy_levels}~(b). The energy spacing in this strong field regime is $E_{\mathcal{MF}}=\mu_\mathrm{B} B_z$, which is of the order of $\sim100..250$~MHz.

The finestructure furthermore yields doublets for neighboring states with $(\Delta M_S, \Delta m_l)= (\pm 1,\mp 2)$, sketched as red and blue lines in \fref{fig:magnetic_field:energy_levels}~(b). The only singlet states with $m_l\neq 0$ are the manifolds $(M_S,m_l) = (N/2,1)$ and $(M_S,m_l)=(-N/2,-1)$. Hence we concentrate on the $(M_S,m_l) = (N/2,1)$ manifold, which can be well addressed during the Rydberg excitation process.
The magnetic field yields increasing decoupling of the $(M_S,m_l) = (N/2,1)$ manifold from the $(M_S,m_l) = (N/2,-1)$ manifold with increasing magnetic field strength. We study this decoupling in detail, giving the Hamiltonian structure in \aref{isotrope_dip_dip_int:setup} and derive effective interactions in \aref{isotrope_dip_dip_int:blockdiag}. The only coupling of the $(M_S,m_l)= (N/2,1)$ state to other manifolds than $(M_S,m_l) = (N/2,-1)$ is through spin-orbit interactions and can thus be neglected.
%
\subsection{Planar Rydberg aggregates in strong B-fields}
\label{isotrope_dip_dip_int:setup}
%
Here we derive the Hamiltonian for a Rydberg aggregate in an external magnetic field pointing in the z-direction, where the magnetic field shift is much larger than the finestructure. This strong field regime allows a reduction of the electronic Hilbert space to a single spin manifold. All atoms are assumed to be located within a plane, as in all cases considered here, with quantization axis perpendicular to that plane. Then, the $m_l=0$ manifold is already completely decoupled \cite{moebius:cradle}. We are interested in $m_l=-1,1$ and thus neglect the $m_l=0$ manifold in the following. 
We use the aggregate states $\ket{\pi_{k}}$ as in the main article and denote with $\ket{-1}, \ket{1}$ the states of angular momentum magnetic quantum number $m_l=-1,1$.
The combined basis is
$\ket{\pi_{k},m_l} = \ket{\pi_{k}}\otimes \ket{m_l} = \ket{s...(p,m_l)...s}$. We consider two Hilbert spaces: The ``pure'' aggregate space, $\spaces$, spanned by the basis $\basiss := \{\ket{\pi_{k}}\}_{k=1}^N $ and the space including the angular momentum magnetic quantum numbers, $\spacel$, spanned by $\basisl := \{\ket{-1},\ket{1}\}\otimes\basiss$.

The dipole-dipole interaction Hamiltonian with magnetic field shift for \emph{fixed} $m_l=\pm 1$ is given by
\begin{equation}
\hat{H}_{m_l}(\vc{R}) :=\hat{H}^{\rm{el}}(\vc{R}) + m_l E_{\mathcal{MF}}\id[{\spaces}],
\end{equation}
with $\hat{H}^{\rm{el}}(\vc{R})$ the Hamiltonian in \eqref{eq:elechamiltonian} and $\id[{\spaces}]$ the identity operator, acting on states of $\spaces$.

The dipole-dipole transitions from $m_l=1$ to $m_l=-1$ are described by
\begin{equation}
 \hat{W}(\vc{R}) := \frac{d^2}{2}\sum_{p,q=1}^N R_{pq}^{-3}\mathrm{e}^{-2\im \phi_{pq}}\ket{\pi_{p}}\bra{\pi_{q}},
\end{equation}
where $d:=d_{\nu,1;\nu,0}$ is the radial dipole matrix element between $p$ and $s$ for principal quantum number $\nu$, and $R_{pq}=|\bv{R}_{pq}|$ and $\phi_{pq}$ are the modulus and azimuthal angle of the separation $\bv{R}_{pq}$ between atoms $p$ and $q$, 
within the co-ordinate system defining our quantization axes.

We treat now $\hat{H}_{m_l}(\vc{R})$ as unperturbed system and $\hat{W}(\vc{R})$ as perturbation and set up extended Hamiltonians:
\begin{align}
 \hat{\mathcal{H}}_0(\vc{R})&:=\sum_{m_l \in \{-1,1\}} \ket{m_l}\bra{m_l}\otimes\hat{H}_{m_l}(\vc{R}),
\label{eq:H0_big}\\
\hat{\mathcal{W}}(\vc{R})&:=\ket{1}\bra{-1}\otimes \hat{W}(\vc{R}) + \ket{-1}\bra{1}\otimes \hat{W}^{\dagger}(\vc{R}),
\label{eq:W_MF_big}\\
\hat{\mathcal{H}}(\vc{R})&:=\hat{\mathcal{H}}_0(\vc{R}) + \hat{\mathcal{W}}(\vc{R}).
\label{eq:H_MF_big}
\end{align}
The Hamiltonian \eqref{eq:H_MF_big} describes Rydberg aggregates with magnetic field shifts for the $m_l=-1, 1$ states, but neglecting fine structure shifts.

\subsection{Effective interactions from block-diagonalization}
\label{isotrope_dip_dip_int:blockdiag}
%
To see how the mixing of the $m_l=-1,1$ manifolds due to the interaction described by $\hat{\mathcal{W}}(\vc{R})$ is weakened with increasing
magnetic field, we block-diagonalize $\hat{\mathcal{H}}(\vc{R})$, according to van Vleck perturbation theory~\cite{vanVleck:PT} in a  canonical form as outlined by Shavitt et al~\cite{shavitt:vanVleck}. 
We define the projectors
\begin{align}
 \hat{\mathcal{P}}_{m_l}&:=\ket{m_l}\bra{m_l}\otimes \id[{\spaces}]
\end{align}
and denote for every operator $\hat{\mathcal{A}}$ that acts on states in $\spacel$ the diagonal blocks with
\begin{equation}
 \hat{\mathcal{A}}_{D} = \hat{\mathcal{P}}_{-1} \hat{\mathcal{A}} \hat{\mathcal{P}}_{-1} + \hat{\mathcal{P}}_{1} \hat{\mathcal{A}} \hat{\mathcal{P}}_{1}
\end{equation}
and the offdiagonal blocks  with
\begin{equation}
 \hat{\mathcal{A}}_{X} = \hat{\mathcal{P}}_{1} \hat{\mathcal{A}} \hat{\mathcal{P}}_{-1} + \hat{\mathcal{P}}_{-1} \hat{\mathcal{A}} \hat{\mathcal{P}}_{1}.
\end{equation}
The goal is to find a unitary transformation, $\hat{\mathcal{U}}(\vc{R})$, such that
$\mathcal{H}'(\vc{R}) = \hat{\mathcal{U}}^{\dagger}(\vc{R})\hat{\mathcal{H}}(\vc{R})\hat{\mathcal{U}}(\vc{R})$ is blockdiagonal, i.e. $\mathcal{H}'_{X}(\vc{R}) = 0$.
In the following we omit the dependency of the operators on $\vc{R}$ for better readability.

In canonical van Vleck perturbation theory, the transformation is rewritten as $\hat{\mathcal{U}} = \mathrm{e}^{\hat{\mathcal{G}}}$ with the property $\hat{\mathcal{G}} = -\hat{\mathcal{G}}^\dagger$ and $\hat{\mathcal{G}}_{D} = 0$. These conditions lead to the following form of $\hat{\mathcal{G}}$:
\begin{equation}
 \hat{\mathcal{G}} =\ket{1}\bra{-1}\otimes \hat{G} - \ket{-1}\bra{1}\otimes \hat{G}^{\dagger},
\label{eq:G_form}
\end{equation}
with $\hat{G}$ acting on states in $\spaces$. An expansion in orders of the perturbation $\hat{\mathcal{W}}(\vc{R})$ \cite{shavitt:vanVleck} leads to equations for $\hat{\mathcal{G}}_{(i)}$, $i$ labeling different orders:
\begin{align}
 [\hat{\mathcal{H}}_0,\hat{\mathcal{G}}_{(1)}] &= -\hat{\mathcal{W}}
\label{eq:det_Gl1}\\
[\hat{\mathcal{H}}_0,\hat{\mathcal{G}}_{(2)}] &= 0
\label{eq:det_Gl2}\\
[\hat{\mathcal{H}}_0,\hat{\mathcal{G}}_{(3)}] &= -1/3[[\hat{\mathcal{W}},\hat{\mathcal{G}}_{(1)}],\hat{\mathcal{G}}_{(1)}]
\label{eq:det_Gl3}\\
&\ \ \vdots\nonumber
\end{align}
Eqns.~\eqref{eq:det_Gl1} - \eqref{eq:det_Gl3} already exploit that in our case $\hat{\mathcal{W}}_{D} = 0$. Using \eqref{eq:G_form} and \eqref{eq:H0_big} to evaluate the lhs.~of these equations, we get
\begin{equation}
 [\hat{\mathcal{H}}_0,\hat{\mathcal{G}}_{(i)}] = \ket{1}\bra{-1}\otimes( [\hat{H},\hat{G}_{(i)}] + 2E_{\mathcal{MF}}\hat{G}_{(i)}) +\mbox{c.c.} .
\end{equation}
Evaluating also the rhs of \eqref{eq:det_Gl1} - \eqref{eq:det_Gl3}, we obtain formulas for the $\hat{G}_{(i)}$:
\begin{align}
 [\hat{H},\hat{G}_{(1)}] &= -2E_{\mathcal{MF}}\hat{G}_{(1)} - \hat{W},
\label{eq:det_Gs1}\\
[\hat{H},\hat{G}_{(2)}] &= -2E_{\mathcal{MF}}\hat{G}_{(2)},
\label{eq:det_Gs2}\\
[\hat{H},\hat{G}_{(3)}] &=
-2E_{\mathcal{MF}}\hat{G}_{(3)}+\hat{F}(\hat{W},\hat{G}_{(1)}),
\label{eq:det_Gs3}
\end{align}
where $\hat{F}(\hat{W},\hat{G}_{(1)}) = 1/3(\hat{W}[\hat{G}_{(1)}^{\dagger},\hat{G}_{(1)}]+\{\hat{W},\hat{G}_{(1)}\hat{G}_{(1)}^{\dagger}\} + 2\hat{G}_{(1)} \hat{W}^{\dagger} \hat{G}_{(1)})$. The curly brackets denote the anticommutator.
All equations \eqref{eq:det_Gs1} -  \eqref{eq:det_Gs3} are of the form 
\begin{equation}
 \hat{X}~=~a[\hat{Y},\hat{X}]~+~b\hat{Z},
\label{eq:comm_eq_general}
\end{equation}
with constants $a=(-2E_{\mathcal{MF}})^{-1},b =  (\mp2E_{\mathcal{MF}})^{-1}$ and operators $\hat{X} = \hat{G}_{(i)},\hat{Y} = \hat{H},\hat{Z} \in \{\hat{W}, 0, \hat{F}\}$. The solution to this equation is 
\begin{equation}
 \hat{X}~=~b\sum_{k=0}^{\infty}a^k[\hat{Y},\hat{Z}]_{k},
\label{eq:solution_comm_eq_general}
\end{equation}
with $[\hat{Y},\hat{Z}]_{k} = [\hat{Y},[\hat{Y},\hat{Z}]_{k-1}]$ and $[\hat{Y},\hat{Z}]_{0} = \hat{Z}$.
The expansion of the block-diagonalization is given by
\begin{align}
\hat{\mathcal{H}}'_{(0)} &= \hat{\mathcal{H}}_0,
\label{eq:Hprime0}\\
\hat{\mathcal{H}}'_{(1)} &= 0,
\label{eq:Hprime1}\\
\hat{\mathcal{H}}'_{(2)} &= 1/2[\hat{\mathcal{W}},\hat{\mathcal{G}}_{(1)}],
\label{eq:Hprime2}\\
\hat{\mathcal{H}}'_{(3)} &= 1/2[\hat{\mathcal{W}},\hat{\mathcal{G}}_{(2)}],
\label{eq:Hprime3}\\
\begin{split}
\hat{\mathcal{H}}'_{(4)} &= 1/2[\hat{\mathcal{W}},\hat{\mathcal{G}}_{(3)}]\\
&-1/24[[[\hat{\mathcal{W}},\hat{\mathcal{G}}_{(1)}],\hat{\mathcal{G}}_{(1)}],\hat{\mathcal{G}}_{(1)}].
\end{split}
\label{eq:Hprime4}
\end{align}
The first correction of the blockdiagonalization procedure is \eqref{eq:Hprime2}. Using solution \eqref{eq:solution_comm_eq_general} for equation \eqref{eq:det_Gs2}, we see that $\hat{G}_{(2)} = 0$ and so $\hat{\mathcal{H}}'_{(3)} = 0$. The second contribution is thus from $\hat{\mathcal{H}}'_{(4)}$, the leading order of which is $\mathcal{O}(E_{\mathcal{MF}}^{-3})$, thus we do not consider it further. We now concentrate on the Hamiltonian for states with $m_l=1$,  $\hat{H}':=\bra{1}\hat{\mathcal{H}}'\ket{1}$. Using \eqref{eq:W_MF_big}, \eqref{eq:G_form} and \eqref{eq:solution_comm_eq_general} in \eqref{eq:Hprime2}, we can evaluate the corresponding first correction term
\begin{align}
\hat{H}'_{(2)} &= (4E_{\mathcal{MF}})^{-1}\sum_{k=0}^\infty (-2E_{\mathcal{MF}})^{-k}[\hat{H},\hat{W}]_{k}\hat{W}^{\dagger} + \mbox{c.c.}.
\label{eq:Hprimeml1_2}
\end{align}
Let us define the maximum dipole-dipole interaction element of $\hat{H}$, $\Eddmax:=d^2/(6\Rmin^3)$, with $\Rmin:=\min_{p,q: p\neq q}R_{pq}$. To assess if \eqref{eq:Hprimeml1_2} is small compared to $\Eddmax$, as required for $\hat{H}' \approx \hat{H}_{m_l=1}$, we introduce rescaled quantities, $\hat{X} := \Eddmax \hat{\tilde{X}}$, with $\hat{X} \in \{ \hat{H}, \hat{H}', \hat{W}  \}$. 
This yields $|\tilde{H}_{pq}| \leq 1$ and $|\tilde{W}_{pq}| \leq 3$. Introducing further a decoupling number, $\alpha:=\Eddmax/(2E_\mathcal{MF})$ and setting $E_{\mathcal{MF}}$ as our zero of energy, we find up to $\alpha^2$ the following effective interaction Hamiltonian for $m_l=+1$:
\begin{equation}
\begin{split}
 \hat{\tilde{H}}' \approx \hat{\tilde{H}} + \dfrac{\alpha}{2}\{\hat{\tilde{W}},\hat{\tilde{W}}^\dagger\}
+\alpha^2 \hat{\tilde{W}}\hat{\tilde{H}}\hat{\tilde{W}}^\dagger
-\dfrac{\alpha^2}{2}\{\hat{\tilde{H}},\hat{\tilde{W}}\hat{\tilde{W}}^\dagger\}
\end{split}
\label{eq:Hprimeml1_2_rescaled}
\end{equation}
Note that $\hat{\tilde{H}}'$ is measured in units of $\Eddmax$. The leading order $\hat{\tilde{H}}$ of the effective Hamiltonian \bref{eq:Hprimeml1_2_rescaled} corresponds to our original isotropic model \bref{eq:elechamiltonian} and the additional terms can be explicitly constructed to allow for small anisotropic corrections.
\begin{figure}[hbt]
\centering
\epsfig{file={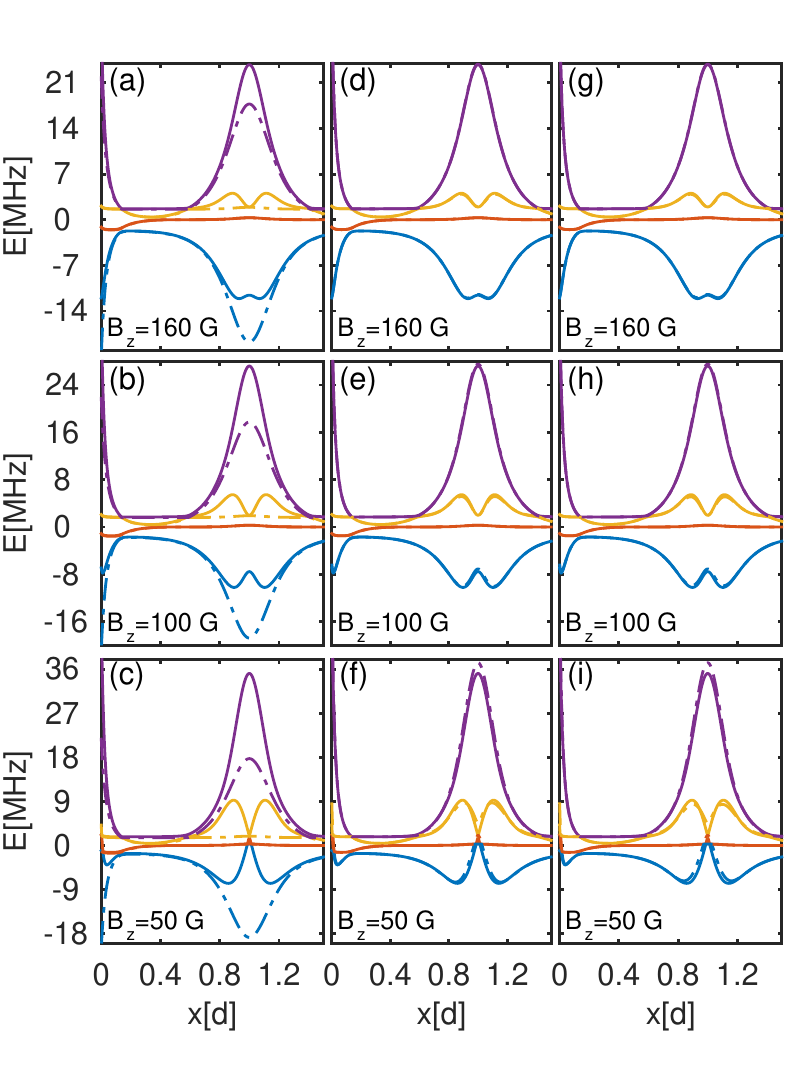},width=\ncolumnwidth} 
\caption{\label{fig:N4_compare_interactions}
(color online) Comparison of energy spectra from different approximations of the complete Hamiltionian, for a double dimer as sketched in \fref{system_sketch_4}~(a). We use $\nu=80$, which yields a finestructure splitting of $\Delta E_{FS}=0.15$~MHz~\cite{haroche:li_finesplitting}. Further parameters are $\ p=0, \ d=40~\mu$m, $a_1=8~\mu$m and $a_2=19~\mu$m. The atoms 2 and 3 are fixed, whereas the positions of atoms 0 and 1 are parameterized, such that $x_0=-x$ for atom~0 and $x_1=a_1+x$ for atom~1. We compare the complete model (solid lines) with different approximate models (dashed dotted lines) in different columns. 
The approximate models are: (a-c) purely isotropic model used for the main results in this paper, with Hamiltonian given in \eqref{eq:elechamiltonian}. (d-f) corrected model with effective Hamiltonian \eqref{eq:Hprimeml1_2_rescaled} up to first order in $\alpha$. (g-i) same as (d-f), but using \eqref{eq:Hprimeml1_2_rescaled} up to second order in $\alpha$.
We consider three different magnetic field strengths:
$B_z=50$~G in the lower row, $B_z=100$~G in the adjacent row and $B_z=160$~G in the upper row. The energy of infinitely separated atoms is set to zero.
}
\end{figure}
%
\subsection{Comparison between isotropic, effective and complete Hamiltonian}
%
In this section we assess how good the isotropic or the effective Hamiltonian (both without spin degrees of freedom) approximate the complete Hamiltonian, which includes spin-orbit coupling in addition to the magnetic field.

We denote the space of the electron spin states of the $i$th-atom with $\spacepinsmaller{i}$, the basis of which is spanned by $\basisspinsmaller{i}~=~\{\ket{-1/2}_{(i)},~\ket{1/2}_{(i)}\}$. The states $\ket{-1/2}_{(i)}$ denote downwards oriented electron spin and $\ket{1/2}_{(i)}$ upwards oriented electron spin for the $i$th-atom.
The space of all $N$ electron spins is then given by $\spaceallspins = \otimes_{i=1}^{N}\spacepinsmaller{i}$, with the product basis $\basissallspins =\otimes_{i=1}^{N} \basisspinsmaller{i}$.\newline
The spin-orbit interaction destroys the decoupling of the $m_l=0$ states, such that we have to redefine some quantities of \aref{isotrope_dip_dip_int:setup}. The space $\spacel$ is now spanned by $\basisl:=\{\ket{-1},\ket{0},\ket{1}\}\otimes \basiss$, with $\ket{m_l}$ the states of the quantum number $m_l \in \{-1,0,1\}$. The Hamiltonian for the $m_l=0$ states is given by
\begin{equation}
 \hat{H}_{0}(\vc{R}):=-2\hat{H}_{\rm{dd}}(\vc{R}) + \hat{H}_{\rm{vdw}}(\vc{R}),
\end{equation}
with $\hat{H}_{\rm{dd}}(\vc{R})$ the resonant dipole-dipole Hamiltonian in \eqref{eq:elechamiltonian-dd} and $\hat{H}_{\rm{vdw}}(\vc{R})$ the non-resonant van-der-Waals Hamiltonian in \eqref{eq:elechamiltonian-vdw}. Note that $\hat{H}_{0}(\vc{R})$ experiences no magnetic field shift. We
redefine $\hat{\mathcal{H}}_{0}(\vc{R})$ from \eqref{eq:H0_big}, such that it includes the $m_l=0$ states:
\begin{equation}
 \hat{\mathcal{H}}_0(\vc{R}):=\sum_{m_l \in \{-1,0,1\}} \ket{m_l}\bra{m_l}\otimes\hat{H}_{m_l}(\vc{R}).
\label{eq:H0_big_redef}
\end{equation}
The Hamiltonian in \eqref{eq:H_MF_big}, $\hat{\mathcal{H}}(\vc{R})$, is now calculated with the redefined $\hat{\mathcal{H}}_0(\vc{R})$.

With these definitions, we can span the complete space $\spacefull := \spacel\otimes \spaceallspins$, describing the orientation of the p-states and the spins of the electrons together. The product basis, where spin and orbital angular momentum of the p-states are not combined to a total angular momentum, is then given by $B_{ls}[\spacefull]:=\basisl \otimes \basissallspins$.

The Hamiltonian $\mathcal{H}(\vc{R})$ in \eqref{eq:H_MF_big} includes the magnetic field shifts for the orbital angular momentum only. The magnetic field shift for the spins is described by
\begin{equation}
 \mathcal{H}_{\mathcal{MF-S}}:=E_{\mathcal{MF}}\sum_{\ket{\vc{M}_S} \in \basissallspins}M_S(\vc{M}_S)\ket{\vc{M}_S}\bra{\vc{M}_S}
\end{equation}
The dipole-dipole interactions together with the total magnetic field shift is then given by
\begin{equation}
 \hat{\stylecs{H}}_{\mathcal{DD}+\mathcal{MF}}(\vc{R}) := \mathcal{H}(\vc{R})\otimes \mathcal{H}_{\mathcal{MF-S}}.
\end{equation}

To set up the spin-orbit interaction Hamiltonian in a simple way, it is useful to employ yet another basis. First we define the spin spaces $\spacepinsmaller{\neq i }$, which describe all spins except those of the 
the $i$th-atom, $\spacepinsmaller{\neq i }:=\otimes_{k\neq i}^{N}\spacepinsmaller{k}$. Their product basis $\basisspinsmaller{\neq i}$ is spanned by $\basisspinsmaller{\neq i}=\otimes_{k\neq i}^{N} \basisspinsmaller{k}$.
The spin-orbit coupling yields a total angular momentum, $\hat{\vc{J}} := \hat{\vc{L}} + \hat{\vc{S}}$ per atom. The pair $(j,m_j)$ are the quantum numbers of $\hat{\vc{J}}$, with $j \in \{ 1/2, 3/2\}$ and $m_j \in M_j:=\{ - |j|, - |j|+1, \dots, |j|\}$.
The result of the spin-orbit coupling is the finestructure splitting $\Delta E_{FS}$, between p-states with $j=3/2$ and p-states with $j=1/2$. To write down the spin-orbit Hamiltonian in its eigenbasis, we first introduce aggregate states which include the spin of the p-states, $\ket{\pi_k,j,m_j}:=\ket{s\dots (p,j,m_j)...s}$. We now define spaces $\spacefullj{j} \subset \spacefull$, which we span with the basis $\basissfullj{j}:=\otimes_{k=1}^N \{\ket{\pi_k,j,m_j}\}_{m_j \in M_j}\otimes \basisspinsmaller{\neq k}$. The orthogonal sum of both '$j$-spaces' spans the complete space, $\spacefull=\spacefullj{1/2}\oplus \spacefullj{3/2}$. This yields the eigenbasis of the spin-orbit Hamiltonian,
$B_{j,mj}[\spacefull]:=\basissfullj{1/2}\cup\basissfullj{3/2}$. Introducing the unitary transformation $\hat{\stylecs{U}}$, which performs the basis transformation from $B_{j,mj}[\spacefull]$ to $B_{ls}[\spacefull]$, the spin-orbit Hamiltonian in the basis $B_{ls}[\spacefull]$ is given by
\begin{equation}
 \hat{\stylecs{H}}_{\mathcal{SO}} = \Delta E_{FS}\hat{\stylecs{U}}\biggl(\nullop[\spacefullj{1/2}] \oplus \id[\spacefullj{3/2}]\biggr)\hat{\stylecs{U}}^\dagger,
\label{eq:H_SO_complete}
\end{equation}
where $\nullop[\spacefullj{1/2}]$ is the null-operator acting on elements in $\spacefullj{1/2}$ and transforming them into its zero. Note that we thus shift the origin of energy to the $j=1/2$ manifold. The complete Hamiltonian is given by
\begin{equation}
 \hat{\stylecs{H}}(\vc{R}):=\hat{\stylecs{H}}_{\mathcal{DD}+\mathcal{MF}}(\vc{R}) + \hat{\stylecs{H}}_{\mathcal{SO}}.
\label{eq:H_complete}
\end{equation}

We compare the three different Hamiltonians in \eqref{eq:elechamiltonian}, \eqref{eq:Hprimeml1_2_rescaled} and \eqref{eq:H_complete} by using them to calculate the eigenenergies for a four atom system with a symmetric configuration ($p=0$), as sketched in \fref{system_sketch_4}~(a). 
We show for \eqref{eq:H_complete} only the $(M_S,m_l) = (N/2,1)$ manifold, which is the one, we propose to work with.
The positions of atoms~(2,3) are fixed, whereas the positions of atoms~0 and 1 are parameterized as $x_0=-x$ and $x_1=a_1+x$. The eigenenergies of all three Hamiltonians are plotted as a function of the co-ordinate $x$ in \fref{fig:N4_compare_interactions}. The isotropic model in \eqref{eq:elechamiltonian} approximates \eqref{eq:H_complete} well for all locations crucial in our simulations, as shown in \fref{fig:N4_compare_interactions}~(a-c). Crucial for the simulations are configurations with small $x$ values, where the atoms are accelerated due to the interactions, and in the neighborhood of the conical intersection. The agreement is not good for the equidistant linear trimer configuration $x = d$. The excitation is there mostly delocalized and the phase of the dipole-dipole interaction plays a role.
The effective Hamiltonian in \eqref{eq:Hprimeml1_2_rescaled} approximates the complete one for this configuration very well, as shown in \fref{fig:N4_compare_interactions}~(d-g). It appears that the Hamiltonian of order $\alpha$ in \eqref{eq:Hprimeml1_2_rescaled} approximates \eqref{eq:H_complete} better than the order $\alpha^2$ version. This may be since \eqref{eq:Hprimeml1_2_rescaled} does not take the spin-orbit coupling into account and the finestructure is of the order of the $\alpha^2$ corrections. A better description beyond the $\alpha$ correction would then require a block-diagonalization, which explicitly includes spin-orbit coupling.

As expected increasing the magnetic field strength improves the decoupling of the $(M_S,m_l) = (N/2,1)$ manifold. This results in a better agreement between the reduced models and the complete model for higher field strengths.


\begin{thebibliography}{54}
\bibitem[{foo({\natexlab{a}})}]{footnote:Umidadj}
\bibinfo{note}{Note that $\ket{\sub{\varphi}{adj}}$ is called
  $\ket{\sub{\varphi}{mid}}$ in \rref{leonhardt:switch}.}
\bibitem[{dis()}]{discussion:CI_crossing}
\bibinfo{note}{Note that in the energy spectra for the four (seven) atom case,
  there is one crossing (are three crossings) of eigenenergies already before
  the wavepacket splitting takes place. These are formally also CIs. However
  they simply reflect the fact that two eigenvalues of excitons localized on
  spatially far separated sites cross. These would also cross if our two chains
  were not coupled. Kinematically, the velocity $\dot{\bv{R}}$ is constrained
  such that these earlier CIs are always exactly hit, causing a diabatic
  transition between surfaces. Also in our simulations, all trajectories safely
  jump to the correct surface at these points.}
\bibitem[{foo({\natexlab{b}})}]{footnote:asymmforce}
\bibinfo{note}{This counter-intuitive fact is due to the excitation
  distribution among the three atoms on the middle energy surface. Here the
  excitation resides predominantly on the atom farthest away, since forces are
  determined by a combination of excitation probability and relative distance,
  in this case this farthest atom experiences the largest force.}
\bibitem[{foo({\natexlab{d}})}]{footnote:purity}
\bibinfo{note}{The purity is defined by $P=\rm{Tr[}\hat{\sigma}^2\rm{]}$, where
  $\hat{\sigma}=\sum_{n,m}\sigma_{n,m}\ket{\pi_n}\bra{\pi_m}$ is the reduced
  electronic density matrix, which describes the electronic state of the
  system, after tracing over the atomic positions. The matrix elements are
  given by $\sigma_{nm}=\langle c_n^* c_m \rangle_{T}$ for the surface hopping
  method. The brackets denotes the trajectory average.}
\end{thebibliography}
\end{document}